\documentclass[12pt]{article}
\usepackage[hmargin=1in,vmargin=1in]{geometry}
\usepackage{setspace}
\usepackage{algorithm}% http://ctan.org/pkg/algorithm
\usepackage{algpseudocode}
\usepackage[utf8]{inputenc}

\usepackage{url}
\usepackage{graphicx}
\usepackage{xcolor,yfonts,eufrak}
\usepackage{blkarray}
\usepackage{amsmath,amssymb,amsthm,amsfonts,bbm,bm,color,setspace,graphics,indentfirst,caption,multicol,multirow,longtable,tabu,verbatim,enumerate,footnote,multirow}
\usepackage{tikz}
\usepackage{pifont}
\usetikzlibrary{shapes,decorations,arrows,calc,arrows.meta,fit,positioning}
\tikzset{
    -Latex,auto,node distance =1 cm and 1 cm,semithick,
    state/.style ={ellipse, draw, minimum width = 0.7 cm},
    point/.style = {circle, draw, inner sep=0.04cm,fill,node contents={}},
    bidirected/.style={Latex-Latex,dashed},
    el/.style = {inner sep=2pt, align=left, sloped}
}
\usepackage{arydshln}
\usepackage{ulem}
\usepackage[sort,round]{natbib}
\title{A Bayesian Joint Model for  Compositional Mediation Effect Selection in Microbiome Data}

\author{Jingyan Fu\thanks{Department of Statistics, Rice University, Houston TX 77005, USA},  Matthew D. Koslovsky\thanks{Department of Statistics, Colorado State University, Fort Collins, CO 80523, USA}, Andreas M. Neophytou\thanks{Department of Environmental \& Radiological Health Sciences, Colorado State University, Fort Collins, CO 80523, USA}, and Marina Vannucci$^*$}
\date{ }

\begin{document}

\maketitle

\begin{abstract}
Analyzing multivariate count data generated by high-throughput sequencing
technology in microbiome research studies is challenging due to the high-dimensional and compositional structure of the data and overdispersion. 
In practice, researchers are often interested in investigating how the microbiome may mediate the relation between an assigned treatment and an observed phenotypic response. Existing approaches designed for compositional mediation analysis are unable to simultaneously determine the presence of direct effects, \textcolor{black}{relative} indirect effects, and overall indirect effects, while   quantifying their uncertainty. We propose a formulation of a Bayesian joint model for compositional data that allows for the identification, estimation, and uncertainty quantification of various causal estimands in high-dimensional mediation analysis. We conduct simulation studies and compare our method's mediation effects selection performance with existing methods.  Finally, we apply our method to a benchmark data set investigating the sub-therapeutic antibiotic treatment effect on body weight in early-life mice. 
\end{abstract}


\section{Introduction}
The human microbiome is the collection of  micro-organisms (e.g., bacteria, archaea, viruses, fungi) that live on and inside of our bodies. A major research question in human microbiome studies is the feasibility of designing interventions that modify the composition of the microbiome to promote health and cure disease. Methodological developments designed to address this research question have taken on various forms and are challenged by the compositional structure, high-dimensionality, overdispersion, and zero-inflation characteristic of microbial count data. Examples of recent developments include sparsity-induced univariate and multivariate count regression models to identify exposures that characterize the composition of the microbiome \citep{xu2015assessment, zhang2020nbzimm,  jiang2021bayesian,  liu2021statistical,zhang2017regression, wadsworth2017integrative,chen2013variable, koslovsky2020microbvs,koslovsky2023bayesian}, compositional regression models to predict biological, genetic, clinical, or experimental conditions using microbial abundance data \citep{lin2014variable}, and joint models for simultaneous inference of these relations \citep{koslovsky2020}, among others. 

Several clinical studies have hypothesized that the microbiome may mediate the relation between an assigned treatment (e.g., diet) and an observed phenotypic response (e.g., body mass index). The total effect of the treatment on the outcome is then comprised of a direct effect (not through the microbiome) and an indirect effect through its relation with the compositional mediators, both of which may be confounded by other covariates. Hypothesis testing and regularization techniques have been proposed to test and identify mediation effects of the microbiome. For example, \cite{Zhang2018} designed a distance-based approach which incorporates prior structural information of the microbial data, such as evolutionary relations, and uses a robust, permutation-based approach for simultaneous inference on multiple distances. This approach estimates an overall mediation effect for the microbiome but cannot estimate  mediation effects for each taxon and does not allow for additional covariates in the model. \cite{sohn2017} assumed a linear log-contrast model to model the relation between potential mediators and the outcome and applied a debiased regularization procedure for estimation to produce both overall and component-wise mediation effect estimates while allowing for additional covariates. \cite{zhang2019} took a similar approach as \cite{sohn2017} but applied isometric log-ratio transformations, often referred to as balances \citep{egozcue2003}, to model the relation between the microbial taxa and the outcome. Thereafter, inference for \textcolor{black}{relative} indirect effects is performed using a joint significance test with a focus on pre-specified taxa. \textcolor{black}{Here, following these authors, 
we refer to the indirect or mediation effects for each taxon as relative indirect effects, to reflect the relative nature of the information captured in each balance (see section \ref{jointmodel} for more details).}    \cite{zhang2020} extended the work of \cite{zhang2019} via a closed testing-based selection procedure to identify individual taxa that mediate the relation between the exposure and phenotypic outcome. \citet{wang2019} proposed a two-stage regularized estimation approach for high-dimensional compositional mediation analysis, which uses a Dirichlet regression model to characterize the relation between treatment and the microbial data while simultaneously investigating potential interaction terms. Similar to \cite{sohn2017}, this model identifies \textcolor{black}{relative} and overall mediation effects in addition to accommodating other covariates and interaction terms. \cite{song2020bayesian} and \cite{song2021bayesian} demonstrate the benefits of a Bayesian approach for exploratory high-dimensional mediation analysis using various types of shrinkage priors to identify active mediators while simultaneously quantifying model uncertainty. However, these approaches are designed for a high-dimensional set of \textit{continuous} mediators which is not suitable for the compositional structure of microbiome data.

%To investigate a high-dimensional set of \textit{continuous} mediators in a Bayesian framework, \cite{song2020bayesian} employed continuous spike-and-slab priors to induce sparsity which shrink excluded covariates' regression coefficients towards zero but do not explicitly remove the covariate from the model. \cite{song2021bayesian} take a similar  used product threshold Gaussian priors for selection, which serve as an alternative to shrinkage priors are are equivalent to non-local priors.

%\textcolor{purple}{In the above section, add some comments regarding what the methods are able to estimate, e.g., overall mediation and marginal, marginal only, etc. Also, comment on whether or not the methods allow for additional covariates in the model and any limitations with estimating model uncertainty (e.g., do they estimate CIs for the marginal effects but not the overall.) This will help us demonstrate the contribution of our approach. }

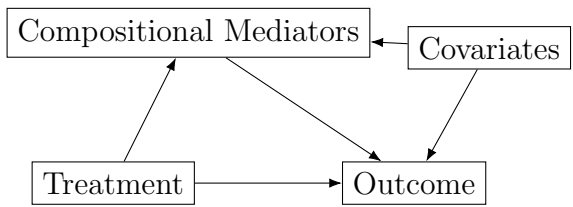
\begin{figure}[hbt]
\centering
\vspace{5mm}
\begin{tikzpicture}
    \node (1) [state,rectangle] (T) at (0,0) {Treatment};
    \node (2) [state,rectangle] (Y) at (4,0) {Outcome};
    \node (3) [state,rectangle, xshift = -1cm] (M) at (2,2) {Compositional  Mediators};
    \node (4) [state,rectangle] (C) at (5,1.8) {Covariates};

    \path (T) edge (Y);
    \path (T) edge (M);
    \path (M) edge (Y);
    \path (C) edge (M);
    \path (C) edge (Y);
\end{tikzpicture}
\vspace{5mm}
\caption{Causal directed acyclic graph of the assumed compositional mediation framework with auxiliary covariates (measured or unmeasured) in both levels of the model.} \label{fig:F1}
\end{figure}

In this paper we build upon the approach of \cite{koslovsky2020} and recast their joint model for compositional microbiome data into the causal framework represented in Figure~\ref{fig:F1}, which allows for the identification of mediation effects under the assumption  of a randomized treatment. \textcolor{black}{
 Compared to two-step approaches, which first model the relation between microbial abundances and a set of covariates and then regress a phenotypic outcome on the estimated relative abundances obtained in the first step, \cite{koslovsky2020} propose jointly modeling phenotypic outcomes and microbial abundances, which directly accommodates uncertainty in the abundance estimates. Notably, their approach is related to the broad class of methods that makes distributional assumptions for covariates to reduce inferential biases \citep{carroll2006measurement, tadesse2005bayesian}.  In simulation, they show that this results in improved selection, estimation, and predictive performance.} To accommodate overdispersion, the microbial abundance data are assumed to follow a Dirichlet-multinomial distribution, given the treatment assignment and a set of observed covariates. \textcolor{black}{A compositional linear regression model relates the relative abundances, which represent the proportion of each microbe in the microbial sample, to the outcome.} We show how the use of \textit{discrete} spike-and-slab priors for regression coefficients, which explicitly place a point mass at zero for excluded terms, provides direct inference on the presence of overall and \textcolor{black}{relative} mediation effects, treatment effects, and potential confounders. By using a fully Bayesian approach for inference, our method inherently quantifies uncertainty for each term in the model and functions thereof. As such it provides a more comprehensive approach for compositional mediation analysis compared to existing approaches.  We demonstrate our method's performance versus comparative approaches on simulated data, provide recommendations for hyperparameter specifications and estimation of causal estimands, and apply our model to a benchmark study investigating the meditation effects of the gut \textcolor{black}{microbiome} on the relation between sub-therapeutic antibiotic treatment and body weight in early-life mice \citep{Shulfer2019}.

In section 2, we first present the Bayesian joint model for compositional mediation analysis and then describe inference on direct and indirect effects following specification of the causal assumptions. 
In section 3, we demonstrate our method's performance in various simulated settings and compare the results to existing methods.  In section 4, we  apply our model to the benchmark study.  We conclude with final remarks in section 5.

\section{Methods}
Let $y_i$ denote the observed continuous outcome of subject $i = 1,\dots,n$ and $t_i\in \{0,1\}$ the assigned treatment, with $t_i = 1$ if subject $i$ received the treatment and $t_i = 0$ otherwise. Furthermore, let $\boldsymbol{z}_i = (z_{i1},...,z_{iJ})^{\prime}$ indicate a $J$-dimensional vector of taxa counts and $\boldsymbol{x}_i = (x_{i1},...,x_{iP})^{\prime}$ a $P$-dimensional vector of observed covariates. We first recast the joint model for compositional microbiome data of \cite{koslovsky2020} into a framework for mediation analysis, where the \textcolor{black}{relative abundances} are treated as potential mediators, and then describe inference on direct and indirect effects following specification of the causal assumptions.

\subsection{Bayesian Joint Model for Mediation Effect Selection}\label{jointmodel}

We adopt a joint model formulation that comprises a linear regression model for the phenotypic outcome and a Dirichlet-multinomial regression model for the compositional taxa. The two models are linked via   balances, calculated based on estimated \textcolor{black}{relative abundances}, that serve as the shared parameters.

{\bf Outcome Model:}
A multiple linear regression model is used to capture the direct effect of the treatment on the outcome, while adjusting for potential mediators and other covariates \textcolor{black}{(including potential confounders of the outcome and mediators)}, as
\begin{eqnarray}
y_i = c_0 + c_1 t_i + \sum_{j=1}^{J-1} \beta_j B(\boldsymbol{\eta}_j,\boldsymbol{\psi}_{i})+ \sum_{p=1}^P \kappa_p x_{ip}  + \epsilon_i, \label{stage2}
\end{eqnarray}
where the balances $B(\boldsymbol{\eta}_j,\boldsymbol{\psi}_{i})$ are a function of the \textcolor{black}{relative abundances} $\boldsymbol{\psi}_i=(\psi_{i1}, \ldots,\psi_{iJ})^{\prime}$, with $\sum_{j=1}^J \psi_{ij} = 1$, as described below. \textcolor{black}{The relative abundances represent the proportion of the microbiome sample that is made up of each microbe.} Regression coefficients $\boldsymbol{\beta} = (\beta_1,\dots,\beta_{J-1})^{\prime}$ represent the balances' effects, $c_0$ the intercept term, and $c_1$ the direct effect of  treatment. Coefficients $\boldsymbol{\kappa} = (\kappa_1, \dots, \kappa_P)^{\prime}$ capture the effects of the covariates, $ \boldsymbol{x}_i$, and $\epsilon_i$ represents the error term.
\textcolor{black}{Spike-and-slab priors \citep{george1997approaches,brown1998multivariate, Vannucci2021} are imposed on the coefficients $\bm{\beta}$ and $\bm{\kappa}$, allowing us to investigate whether the balances and/or covariates are associated with the outcome, respectively.} Specifically,
\begin{eqnarray} 
%c_1 \textcolor{black}{\mid} \tau, \sigma^2 \sim& \hspace{-2.6cm}\tau  N(0,h_{c}\sigma^2) + (1- \tau) \delta_0 (c_1), \nonumber 
 \beta_j \textcolor{black}{\mid} \xi_j, \sigma^2   \sim&  \hspace{-.18cm} \xi_j  N(0,h_{\beta}\sigma^2) + (1- \xi_j) \delta_0 (\beta_j), \mbox{ } j=1,\dots,J-1, \label{bprior} \\
\kappa_p \textcolor{black}{\mid} \nu_p, \sigma^2 \sim& \hspace{-0.6cm} \nu_p   N(0,h_{\kappa}\sigma^2) + (1- \nu_p)   \delta_0 (\kappa_p),\mbox{ } p=1,\dots,P  \nonumber, 
\end{eqnarray}
where $\delta_0(\cdot)$ represents a Dirac delta function, or point mass, at zero. Here, the latent inclusion indicators $\xi_j$ and $\nu_p$ take on values of 0 or 1, where $\xi_j=1$ ($\nu_p=1$) indicates that the corresponding balance (covariate) is included in the model, and 0 otherwise. 
We assume Bernoulli priors on the binary inclusion indicators, with Beta hyperpriors imposed on the inclusion probabilities. This allows the inclusion probabilities to be marginalized out for efficient sampling. We indicate this prior construction as
 $\xi_j \sim  \mbox{Beta-Bernoulli}(a_j,b_j)$ and $\nu_p \sim  \mbox{Beta-Bernoulli}(a_p,b_p)$, where  $a_j$ ($a_p$) and $b_j$ ($b_p$) control the sparsity of the balances (covariates) in the model. To complete the outcome model's formulation, we assume  $c_0, c_1 \sim \mbox{Normal}(0,h_c\sigma^2$) and $\epsilon_i \sim \mbox{Normal}(0,\sigma^2)$, where  $\sigma^2 \sim \mbox{Inverse-Gamma}(a_0,b_0)$ for some $a_0 > 0$ and $b_0 > 0$. 

{\bf Dirichlet-Multinomial Model:} 
The microbial taxa counts are treated as compositional and assumed to follow a multinomial distribution given the \textcolor{black}{relative abundances}  $\boldsymbol{\psi}_i$ (i.e., $\boldsymbol{z}_i \sim \mbox{Multinomial}(\Dot{z_i} \textcolor{black}{\mid} \boldsymbol{\psi}_i)$, where $\Dot{z_i} = \sum_{j=1}^J z_{ij}$). Conjugate priors for $\boldsymbol{\psi}_i$ can be specified as $\boldsymbol{\psi}_i \sim \mbox{Dirichlet}(\boldsymbol{\gamma}_i)$, where $\boldsymbol{\gamma}_i$ is a $J$-dimensional vector of concentration parameters. Note that the distributional assumptions for the taxa counts could take on various forms \citep{zhang2017regression}. We chose a Dirichlet-multinomial (DM) model as it accommodates overdispersion and provides a computationally efficient Markov chain Monte Carlo (MCMC) routine that exploits data augmentation \citep{koslovsky2020}.
A log-linear regression framework can be used to relate the \textcolor{black}{relative abundances}  with the treatment and covariates by introducing $\lambda_{ij} = \log(\gamma_{ij})$ and defining 
\begin{eqnarray}\label{loglinear}
    \lambda_{ij} = \alpha_j + \phi_{j}t_i + \sum_{p=1}^{P} \theta_{jp} x_{ip}. \label{stage1}
\end{eqnarray}
In this formulation, $\alpha_j$ is a taxon-specific intercept term, $\phi_{j}$ is the taxon-specific regression coefficient for treatment, and $\boldsymbol{\theta}_j = (\theta_{j1}, \dots, \theta_{jP})^{\prime}$ are the taxa-specific regression coefficients corresponding to the covariates.   Note that in general the potential covariates included in Equation (Eq.)\ \eqref{stage1} do not have to match those included in Eq.\ \eqref{stage2}.
Similar to the outcome model, influential terms can be identified by imposing spike-and-slab priors on each of the regression coefficients, $\phi_j$ and   ${\theta}_{jp}$, with Gaussian slabs centered at 0 and  variance $r_j^2$. We assume Beta-Bernoulli priors for the latent inclusion indicators, $\varphi_j \sim \mbox{Beta-Bernoulli} (a_v,b_v)$ and $\zeta_{jp} \sim \mbox{Beta-Bernoulli} (a_t,b_t)$, respectively. The prior specification is completed by assuming $\alpha_j \sim$ Normal$(0,\sigma^2_{\alpha})$. 

{\bf Construction of Balances:}
The outcome model and the DM model are linked via   balances, calculated based on \textcolor{black}{relative abundances}, that serve as shared parameters.
Balances are isometric log-ratio transformations,  \textcolor{black}{defined proportionally to the difference in the mean
of the log-transformed abundances between two groups or partitions, and are scale invariant \citep{egozcue2003}.} For a generic balance $k$, the \textcolor{black}{relative abundances} $\boldsymbol{\psi}$ are divided into two non-overlapping partitions, denoted as $\psi_{k+}$ and $\psi_{k-}$, which we represent as a $J$-dimensional vector $\boldsymbol{\eta}_k$. The elements of $\boldsymbol{\eta}_k$ take on values of $1,-1$, or $0$ with indices corresponding to the taxa positions in $\boldsymbol{\psi}$. Specifically, $1$ indicates that the corresponding $\psi_{j}$ belongs to partition $\psi_{k+}$, $-1$
that it belongs to partition $\psi_{k-}$, and $0$ implies it is not in either partition. The balance for a partition is defined as
\begin{eqnarray*}
B(\boldsymbol{\eta}_k,\boldsymbol{\psi}) = \sqrt{\frac{\lvert \psi_{k+} \rvert \lvert \psi_{k-} \rvert}{\lvert \psi_{k+} \rvert + \lvert \psi_{k-} \rvert}} \log\Bigg(\frac{g(\psi_{k+})}{g(\psi_{k-})}\Bigg), \label{Balance}
\end{eqnarray*}
where $\lvert \cdot \rvert$ indicates the dimension of the partition and $g(\cdot)$ the geometric mean. \textcolor{black}{Thus, balances can be seen as a normalized log ratio of the geometric mean of the elements assigned to each partition, and  $\beta_j$ in Eq.\ \ref{stage2} is interpreted at the expected change in $Y$ for a unit increase in the logarithm of the ratio between the geometric mean of the taxa in $\psi_{j+}$  and the taxa in $\psi_{j-}$.}
%As in \cite{koslovsky2020},  
\textcolor{black}{We define the partitions using sequential binary separation \citep{egozcue2005groups}, which we formalize in section \ref{strategies}.  Briefly, given a  vector of relative abundances $\boldsymbol{\psi} = (\psi_1, \psi_2,\dots,\psi_J)$, we generate $J-1$ sequential binary partitions in which the first partition is defined as $\psi_{1+} = \{ \psi_1\}$ and $\psi_{1-} = \{ \psi_2 ,\dots, \psi_J\}$, the second partition is defined as $\psi_{2+} = \{ \psi_2\}$ and $\psi_{2-} = \{ \psi_3 ,\dots, \psi_J\}$, and so on until $\psi_{J-1,+}=\{\psi_{J-1}\}$ and $\psi_{J-1,-}=\{\psi_J\}.$ } It is important to note that prediction performance of the model does not depend on the order in which the partitions are defined using sequential binary separation \citep{koslovsky2020}. Additionally, balances cannot handle observed zero counts and require adjustments
based on assumptions of their occurrence \citep{martin2015bayesian}. To handle zero values for $\boldsymbol{\psi}$, we use a multiplicative replacement strategy in which zero values are replaced with relatively small pseudovalues, and the corresponding probability vector is scaled to sum to one \citep{martin2000zero}. This strategy does not affect the modeling of the relationship between treatment and \textcolor{black}{relative abundances}.

\subsection{Causal Assumptions and Definition of Mediation Effects}\label{causal}
 
We now discuss the assumptions required to identify the direct and indirect causal effects in our modeling approach. We operate under the potential outcomes framework \citep{rubin2005causal}. Within this framework, potential outcomes for each subject exist under any possible treatment value, but the outcome for a subject can only be observed under one treatment value. The potential outcome under the treatment value the subject does not receive (counterfactual treatment) is typically referred to as the counterfactual outcome \citep{hofler2005causal}. The total effect of the treatment on the outcome on the additive scale is the summation of a direct effect and an indirect effect through its relation with the compositional mediators.  One of the key advantages of our approach is that it provides inference on taxon-specific mediation effects as well as an overall mediation effect for the microbiome, in addition to inherently estimating model uncertainty.
\textcolor{black}{ Under the typical stable unit treatment value assumption (i.e., consistency and no interference) \citep{rubin1980randomization,rubin1986statistics}, we assume that: }
\textcolor{black}{ 
\begin{description}
    \item[  \hspace{0.5cm}   Assumption 1.]  $0 < P(T_i = t \textcolor{black}{\mid} \boldsymbol{X}_i=\boldsymbol{x})  < 1$,
    \item[  \hspace{0.5cm}   Assumption 2.]  $0  < P( \bm{B}(\bm{\eta}, \bm{\psi}_i(t))=\bm{b} \textcolor{black}{\mid} T_i=t, \boldsymbol{X}_i=\boldsymbol{x})  <1$, 
    \item[  \hspace{0.5cm}   Assumption 3.] There is no interaction between $T_i$ and $\bm{\psi}_i$,
    \item[  \hspace{0.5cm}   Assumption 4a.] $\{Y_i(t^{\prime}, \bm{b}), \boldsymbol{B}(\bm{\eta}, \boldsymbol{\psi}_i(t))\} \perp T_i \textcolor{black}{\mid} \boldsymbol{X}_i=\boldsymbol{x}$,
    \item[  \hspace{0.5cm}   Assumption 4b.] $Y_i(t^{\prime}, \bm{b}) \perp \boldsymbol{B}(\boldsymbol{\eta}, \boldsymbol{\psi}_i(t)) \textcolor{black}{\mid} T_i = t, \boldsymbol{X}_i=\boldsymbol{x}$,
\end{description} }  
\noindent for all $\boldsymbol{x}$, and where $t,t^{\prime} \in \{0,1\}$ are the assigned and counterfactual  treatments, respectively, $\boldsymbol{B}(\boldsymbol{\eta},\boldsymbol{\psi}_i(t))$ is the corresponding balance set, \textcolor{black}{$\boldsymbol{X}_i$ refers to potential confounders of the outcome and mediators, as well as additional covariates in each model,} and $Y_i(t , \bm{b})$ is the potential outcome when treatment $T_i = t$ and balance $\bm{B}(\bm{\eta}, \bm{\psi}_i(t))=\bm{b}$. Assumptions \textcolor{black}{4a} and \textcolor{black}{4b}, in particular, imply that there are no unmeasured confounders after controlling for the covariates and treatment \citep{Imai2010}. \textcolor{black}{Note that assumptions 1 and 4a are expected to hold for the simulation and application study by definition of the randomized design. In contrast, assumption 4b is an untestable assumption even with a randomly assigned exposure, but we explore the robustness of the method with respect to violations of this assumption in the simulation study. Furthermore, adjustment for post-treatment variables as a subset of $\boldsymbol{X}_i$ that may confound the mediator-outcome relationship can only be made under the additional assumption that they are not induced by the exposure,  tantamount to a cross-world independence assumption \citep{andrews2020insights}. It should also be noted that the assumption of no interaction between the treatment and mediator is not necessary for the identification of direct and indirect effects, though the simulations and applied example in the current study operate under this assumption.  }

Under the assumptions above, we can now define the direct effect of the treatment on the phenotypic response for the $i^{th}$ subject, $\Delta_i$, as
\begin{eqnarray*}
\nonumber \Delta_i &=& E[Y_i(T_i=1,\boldsymbol{B}(\boldsymbol{\eta},\boldsymbol{\psi}_i(T_i))) - Y_i (T_i=0,\boldsymbol{B}(\boldsymbol{\eta},\boldsymbol{\psi}_i(T_i))) \textcolor{black}{\mid \boldsymbol{X}_i = \boldsymbol{x}_i}] \\ 
       &=& c_1(1 - 0) = c_1. \label{direct}
       \end{eqnarray*}
%given the corresponding inclusion indicator $\tau = 1$. 
\textcolor{black}{Note that the direct effect is shared among all subjects as $\Delta_i = \Delta_{i^{\prime}}$, $\forall i \neq {i^{\prime}}$}. \textcolor{black}{The subject-specific  overall indirect effect, $\delta_i$, is then defined as}
\begin{eqnarray*}
\nonumber\delta_i &=& E[Y_i(T_i ,\boldsymbol{B}(\boldsymbol{\eta},\boldsymbol{\psi}_i(T_i=1))) - Y_i( T_i,\boldsymbol{B}(\boldsymbol{\eta},\boldsymbol{\psi}_i(T_i=0)))\textcolor{black}{\mid \boldsymbol{X}_i = \boldsymbol{x}_i}] \\ 
&=& 
 \sum_{j=1}^{J-1} \beta_j  ( E[B(\boldsymbol{\eta}_j, \bm{\psi}_i(T_i = 1, \textcolor{black}{\boldsymbol{X}_i = \boldsymbol{x}_i}))] - E[B(\boldsymbol{\eta}_j, \bm{\psi}_i(T_i = 0, \textcolor{black}{\boldsymbol{X}_i = \boldsymbol{x}_i}))] ), \label{eq:overall}
       \end{eqnarray*}
 
\noindent where
\begin{eqnarray}
\nonumber && E[B(\boldsymbol{\eta}_j, \bm{\psi}_i(T_i = t, \textcolor{black}{\boldsymbol{X}_i = \boldsymbol{x}_i})) ] =   \sqrt{\frac{  J-j}{J-j+1}}\bigg{(} E[\log(\psi_{ij}(T_i = t, \textcolor{black}{\boldsymbol{X}_i = \boldsymbol{x}_i}))  ] -\\ && \hspace{3cm} \frac{1}{J-j}\sum_{k=j+1}^{J}E[\log(\psi_{ij}(T_i = t, \textcolor{black}{\boldsymbol{X}_i = \boldsymbol{x}_i}))]\bigg{)},\\
 &&E[\log(\psi_{ij}(T_i = t, \textcolor{black}{\boldsymbol{X}_i = \boldsymbol{x}_i}))] = \Psi\big{(}\gamma_{ij}(T_i=t,\textcolor{black}{\boldsymbol{X}_i = \boldsymbol{x}_i})\big{)} - \Psi\big{(}\sum_{k=1}^{J}\gamma_{ik}(T_i=t,\textcolor{black}{\boldsymbol{X}_i = \boldsymbol{x}_i})\big{)}  \label{explog} , \nonumber\\
 && \gamma_{ij}(T_i=t,\textcolor{black}{\boldsymbol{X}_i = \boldsymbol{x}_i}) = \exp(\alpha_j + \phi_jT_i + \sum_{p=1}^P \theta_{jp}x_{ip}) \nonumber,
 \end{eqnarray} and $\Psi(\cdot) = \frac{d}{dx} \log(\Gamma(x))$ is the digamma function following \cite{honkela2001nonlinear}, given the corresponding inclusion indicators $\xi_j = \varphi_j = 1$. \textcolor{black}{Note that the subject-specific overall indirect effects vary across subjects, since the estimated relative abundances are a function  of treatment assignment and a set of uniquely observed covariates. When there are no covariates in the DM portion of the   model, the population-level overall indirect effect is the same for each individual.} 

\subsubsection{Strategies for Determining \textcolor{black}{Relative} Mediation Effects}\label{strategies}
\textcolor{black}{Unlike the overall indirect effect, identification of the separate indirect effects is typically subject to the additional assumption of independence between individual mediators (here, individual taxa) conditional on $T_i$ and $\boldsymbol{X}_i$ \citep{imai2013identification, VanderWeele2014}. \cite{kim2019bayesian} have recently proposed an alternative decomposition of  indirect effects of individual mediators from the joint (or overall) indirect effect of multiple mediators using a Bayesian estimation approach, while allowing for  interdependence between mediators.} \textcolor{black}{Their approach relies on a joint distributional assumption for the multiple mediators, similar to the proposed method in which we assume a Dirichlet-multinomial distribution for the multiple mediators. With the proposed approach, the calculation  and interpretation  of the \textcolor{black}{relative} indirect effects for each taxon depends on the order of the taxa when constructing the balances via sequential binary separation,  though the order makes no assumptions about the nature of the causal relationships between individual taxa.} 
To better understand this concept, consider a balance tree structure that is constructed by creating a $(J-1)*J$-dimensional vector $\boldsymbol{\eta}  = (\boldsymbol{\eta}_1,\dots, \boldsymbol{\eta}_{J-1})^{\prime}$, with
\begin{eqnarray*}
\boldsymbol{\eta}_1 &=& (1,-1,-1,\dots,-1,-1) \\
\boldsymbol{\eta}_2 &=& (0, 1,-1,\dots,-1,-1) \\
& & \vdots\\
\boldsymbol{\eta}_{J-1} &=& (0,0,0,\dots, 1,-1),   
\end{eqnarray*}
\textcolor{black}{and calculating the $(J -1)-$dimensional vector of balances $\boldsymbol{B}(\boldsymbol{\eta},\boldsymbol{\psi}) = (B(\boldsymbol{\eta}_1,\boldsymbol{\psi}),\dots,B(\boldsymbol{\eta}_{J-1},\boldsymbol{\psi}))^{\prime}$.}  This implies that the \textcolor{black}{relative} mediation effect of the taxon corresponding to the first element in $\boldsymbol{\eta}_1$, $\delta_{i1}$, given $B(\bm{\eta}_1,\bm{\psi}_i)$, can be defined as
\begin{equation}
\textcolor{black}{ \delta_{i1} = \beta_1\sqrt{\frac{J-1}{J}}\Big(E[\log(\psi_{i1}(T_i=1,\boldsymbol{X}_i =\boldsymbol{x}_i))] - E[\log(\psi_{i1}(T_i=0, \boldsymbol{X}_i =\boldsymbol{x}_i))]\Big), }
 \label{marginal}
\end{equation} 
where \textcolor{black}{$E[\log(\psi_{i1}(T_i=1,\boldsymbol{X}_i =\boldsymbol{x}_i))]$ and $E[\log(\psi_{i1}(T_i=0,\boldsymbol{X}_i =\boldsymbol{x}_i))]$} are evaluated using \textcolor{black}{ Eq.\  \eqref{eq:overall}}, similar to the overall mediation effect. \textcolor{black}{As such, when $\varphi_1^{[1]}$ and $\xi_1^{[1]}$ are both active, or selected, into the model, the taxon corresponding to $\psi_1$ has a significant mediation effect relative to the rest of the taxa.} 

The \textcolor{black}{relative} mediation effects of the remaining taxa are expressed as 
\textcolor{black}{
\begin{eqnarray}
    \nonumber \delta_{i2} = \Big(\beta_2\sqrt{\frac{J-2}{J-1}}
 - \beta_1 \sqrt{\frac{J-1}{J}}\frac{1}{J-1}\Big) \vartheta_{i2}, \\
    \delta_{ij} = \Big(\beta_j\sqrt{\frac{J-j}{J-j+1}}
 - \sum_{k=1}^{j-1} \beta_k \sqrt{\frac{J-k}{J-k+1}}\frac{1}{J-k}\Big)\vartheta_{ij}, \label{Expansionj}\\
    \nonumber \dots \\
    \delta_{iJ} = - \sum_{k=1}^{J-1} \beta_k \sqrt{\frac{J-k}{J-k+1}}\frac{1}{J-k}\vartheta_{iJ}, \nonumber
\end{eqnarray} }
where \textcolor{black}{$\vartheta_{ij}$ =  $E[\log(\psi_{ij}(T_i=1,\boldsymbol{X}_i =\boldsymbol{x}_i)) - E[\log(\psi_{ij}(T_i=0, \boldsymbol{X}_i =\boldsymbol{x}_i))$.} Note that the relative mediation effects depend on the corresponding latent inclusion indicators $\xi_k$ and $\varphi_k$ for $k = 1,\dots,j-1$. \textcolor{black}{Details of these derivations are provided in the Supplementary Material.} \textcolor{black}{Thus, while the \textcolor{black}{relative} mediation effects can be estimated for taxon $j \neq 1$ using our method, we only obtain direct inference regarding the identification of a non-null $\delta_{i1}$ (i.e., $\delta_{i1} \neq 0 $) via its corresponding latent inclusion indicators, $\xi_1$ and $\varphi_1$, for a given ordering of the taxa when constructing the balances. }  This result is an artifact of the compositional structure of the multivariate count data, and is therefore not unique to our modeling approach. \textcolor{black}{Similar to the overall indirect effect, when there are no covariates in the DM portion of the the model, the population-level relative indirect effects are the same for each individual.} 

Given the  definitions described above, we put forward three different strategies to  determine active \textcolor{black}{relative} mediation effects for each taxon and later investigate their performance in the simulation study. Direct inference on the presence of a \textcolor{black}{relative} mediation effect for a given taxon via its corresponding latent inclusion indicators is only available if the taxon is assigned to the first index in $\boldsymbol{\eta}_1$. Therefore, one strategy is to run the MCMC algorithm $J$ times with the balances constructed using a different compositional element in the first index on each run. For each run, we identify the $j^{th}$ taxon-specific mediation effect as active if the marginal posterior probabilities of the corresponding inclusion indicators for $\phi_1^{[j]}$ and $\beta_1^{[j]}$ ($\varphi_1^{[j]}$ and $\xi_1^{[j]}$, respectively) are both greater than or equal to 0.5, where the superscript $[j]$ indicates the $j^{th}$ taxon is the $1^{st}$ element in $\boldsymbol{\eta}_1$. In the comparative study performed in section \ref{comparison} below, we refer to this inferential strategy as CMbvs$_1$.

In order to avoid running the MCMC algorithm $J$ times, an alternative strategy for determining \textcolor{black}{relative} mediation effects is to construct 95\% credible intervals for each of the \textcolor{black}{relative} mediation effects determined using \textcolor{black}{Eqs.\ \eqref{marginal} and \eqref{Expansionj}}. Effects are then identified as active if their 95\% credible intervals for $\delta_{ij}$ do not contain zero. While this approach does not perform inference directly on the latent inclusion indicators, model uncertainty is still propagated into the corresponding effects' credible intervals as the MCMC algorithm iterates through different model parameterizations. We refer to this strategy as CMbvs$_2$. A third potential strategy for determining \textcolor{black}{relative} mediation effects (CMbvs$_3$) involves a combination of strategies 1 and 2. First, the model is run once to determine active treatment effects in the DM portion of the model based on the marginal posterior probabilities of inclusion (MPPIs) for $\varphi_j$. Then the model is re-run with the inactive terms removed. \textcolor{black}{Relative} mediation effects are then determined based on the  95\% credible intervals for $\delta_{ij}$. \textcolor{black}{Note that for CMbvs$_2$ and CMbvs$_3$, selection of active relative indirect effects is determined for each unique covariate profile across subjects. If there are no covariates in the DM portion of the model, selection is performed at the population level. }

\subsection{\textcolor{black}{Posterior Sampling and Inference}}
\textcolor{black}{To sample the posterior distribution, we adopt the Metropolis–Hastings (MH) within Gibbs algorithm of \cite{koslovsky2020} that uses a data augmentation approach to sample the \textcolor{black}{relative abundances} $\boldsymbol{\psi}$.} Let $k_{ij}$ represent latent variables,  such that $\psi_{ij} = k_{ij}/\sum_{j=1}^J  k_{ij}$. Thus, $\boldsymbol{z}_i \sim \mbox{Multinomial}(\Dot{z_i} \textcolor{black}{\mid}\boldsymbol{k}_i^{\prime}/\sum_{j=1}^J  k_{ij})$, with $\boldsymbol{k}_i = (k_{i1}, \dots, k_{iJ})^{\prime}$ and $k_{ij} \sim \mbox{Gamma}(\gamma_{ij},1)$. Introducing auxiliary parameters $\boldsymbol{u}= (u_1, \dots, u_n)^{\prime}$,  such that $u_i \textcolor{black}{\mid} \sum_{j=1}^J  k_{ij} \sim$ Gamma($\Dot{z_i},\sum_{j=1}^J  k_{ij}$), results in closed-form Gibbs updates for $u_i$ and $k_{ij}$. 
Inclusion indicators of the spike-and-slab priors and corresponding regression coefficients are updated jointly using an Add-Delete MH algorithm \citep{savitsky2011variable}. \textcolor{black}{A generic iteration of the MCMC algorithm is described in the Supplementary Material, and more details are provided in \cite{koslovsky2020}.}

Given the output of the MCMC algorithm, MPPIs are used to determine the active terms at both levels of the model. The MPPIs for treatment, covariates, and balances are determined by taking the average of their respective inclusion indicators' MCMC samples after burn-in. Generally, a term is selected if its corresponding MPPI $\geq$ 0.50 \citep{barbieri2004optimal}. \textcolor{black}{ One of the strengths of using \textit{discrete} spike-and-slab priors for Bayesian variable selection is that non-active, or excluded, covariates' corresponding regression coefficients are set to 0 and effectively removed from the model. As a result, MCMC samples for regression coefficients with corresponding MPPIs $< 1.0$  will be zero-inflated (e.g., an MPPI of 0.6 for a given covariate would result in selection using a 0.5 threshold, but 40\% of the corresponding regression coefficient's MCMC samples would be equal to zero). While we favor this approach over refitting the model with selected covariates fixed into the model since it fully accommodates model uncertainty, this may result in   skewed credible intervals and shrink posterior estimates towards zero. }

\section{Simulation Study}
The assumptions made in section \ref{causal} imply that there are no unmeasured confounders in the model.
In observational studies, unmeasured confounding  will result in biased exposure effect estimates \citep{fewell2007impact}. In practice, researchers may employ sensitivity analyses assessing the magnitude of these biases \citep{vanderweele2011unmeasured}. In this simulation study, we evaluate the method's ability to successfully identify and estimate mediation effects in various settings, including unmeasured mediator-outcome confounding (i.e., the presence of covariates associated with the mediators and outcome not accounted for in the model) as well as model misspecification.

%(see  Fig.\ \ref{fig:final}} for details.)
%\begin{figure}[hbt]
%\centering
%\begin{tikzpicture}
%    \node (1) [state,rectangle] (T) at (0,0) {Treatment};
%    \node (2) [state,rectangle, xshift =1cm] (Y) at (4,0) {Outcome};
%    \node (3) [state,rectangle] (M) at (2,2) {Compositional Mediators};
%    \node (4) [state,rectangle, xshift = 2.5cm, yshift =1.2cm] (U1) at (5,0.8) {Unmeasured Confounders};
%   % \node (5) [state,rectangle] (U2) at (-2,1.2) {Confounder2};
%    
%    \path (T) edge (Y);
%    \path (U1) edge (Y);
%    \path (U1) edge (M);
%  %  \path (U2) edge (Y);
%  %  \path (U2) edge (M);
%    \path (T) edge (M);
%    \path (M) edge (Y);
%\end{tikzpicture}
%\caption{Causal directed acyclic graph of the assumed compositional mediation framework with unmeasured confounding in both levels of the %model.} \label{fig:final}
%\end{figure} 
\textcolor{black}{
\subsection{Simulated Data} \label{SimulatedData}
We evaluated the proposed model in various simulated scenarios to demonstrate its \textcolor{black}{relative} mediation effect selection \textcolor{black}{and parameter estimation} performance. In all scenarios, \textcolor{black}{relative abundances} were generated from a Dirichlet distribution and log transformed in the outcome model, instead of transformed via balances. \textcolor{black}{Specifically, we generated the continuous outcome with $y_i = c_0 + c_1 t_i + \sum_{j=1}^{J} \beta_{log,j} \log( \psi_{ij}) + \epsilon_i $, where $\beta_{log,j}$ represents the true regression coefficients specified for the log of the $j^{th}$ relative abundance, \textcolor{black}{similar to \citet{zhang2020}}.} Thus, the data generation process does not match the assumptions of our proposed model. In each scenario, we generated multivariate compositional count data for $n$ = 200 observations with $J$ = 50 compositional elements. We simulated the treatment received by each subject, $t_i$, from a Bernoulli distribution with probability 0.5. Let $t_i$ = 1 indicate that subject $i$ received the treatment, and $t_i$ = 0 otherwise. \textcolor{black}{We assumed  $c_0$ = 0, $c_1$ = 1, and $\epsilon_i \sim \mbox{Normal}(0,1)$ when generating the continuous outcome. For each subject, the relative abundances, ${\bm \psi}_i$, were generated similar to Eq.\ \eqref{stage1}, (i.e., $\lambda_{ij} = \alpha_j + \phi_{j}t_i $)}. The taxon-specific intercept terms $\alpha_j$ were generated from a Uniform$(-2,0.5)$, for $j=1,\ldots,50$.  We simulated data in which the first three taxa have active indirect effects with $\boldsymbol{\phi} = (1, 1.2, 1.5, 0, \ldots, 0)^{\prime}$ and $\boldsymbol{\beta}_{log} = (3, -1.5, -1.5, 0, \ldots, 0)^{\prime}$ for the corresponding log transformed \textcolor{black}{relative abundances}. } 

\textcolor{black}{\textcolor{black}{In the first scenario, we assumed both levels of the model were correctly specified with respect to the casual assumptions} (i.e., no model misspecification or unmeasured mediator-outcome confounding). \textcolor{black}{In scenarios 2 and 3, we evaluated how misspecification of the causal assumptions} (i.e., ignoring influential covariates) in the DM and linear portions of the model, respectively, may affect inference. Scenario 4 introduced unmeasured mediator-outcome confounding due to covariates unaccounted for in both levels of the model. \textcolor{black}{To simulate misspecification of the causal assumptions in  scenarios 2, 3, and 4, we included a binary covariate, $U_{1i}$, simulated from a Bernoulli distribution with probability 0.5 and a continuous covariate, $U_{2i}$, simulated from a standard normal distribution in each layer of the model when generating the data as necessary, but ignored these covariates when fitting the models.}   For the purposes of this simulation, the same covariates affected both the mediators and outcome in scenario 4.  \textcolor{black}{ Specifically, for scenarios 2 and 4, we simulated multivariate count data with $\lambda_{ij} = \alpha_j + \phi_jt_i + \nu_{1j}U_{1i} + \nu_{2j}U_{2i}$, where $\nu_{1j}$ and $\nu_{2j}$ are the $j^{th}$ elements of the $J$-dimensional vectors $\boldsymbol{\nu}_{1} = (0.8, 0, 0, 0, 1.2, 0, \dots, 0)^{\prime}$ and $\boldsymbol{\nu}_{2} = (0, 1.2, 0, 0.8, 0, \dots, 0)^{\prime}$, respectively.  In scenarios 3 and 4, we set $y_i = c_0 + c_1 t_i + \sum_{j=1}^{J} \beta_{log,j} \log( \psi_{ij}) + \kappa_{1}U_{1i} + \kappa_{2}U_{2i} + \epsilon_i $, with $\kappa_{1} = \kappa_{2} = 1.2$.  }}
 %For the unmeasured confounders, we assume they have either strong or no effect on each edges presented in Figure \ref{fig:final}, which results in the following 4 scenarios: 
%\begin{enumerate}
 %   \item Confounders contribute to the %generation of both mediators and outcomes; 
  %  \item Confounders affect the value of mediators, but not the outcomes; 
  %  \item Confounders affect the value of outcomes, but not the mediators;
  %  \item There is no unmeasured confounders within the data generating process.
%\end{enumerate}
\textcolor{black}{We further explored the performance of the method  in scenario 1 under different settings including a skewed distribution for the treatment (i.e., $P(T_i=1) = 0.25$) as well as varying sample sizes and numbers of compositional elements (i.e.,  $n=50$ with $J=50$ and $J=100$). \textcolor{black}{Additionally, we evaluated the models in a setting motivated by the application data with 36 observations, 36 compositional elements, and effect sizes of $\boldsymbol{\phi} = (0.7,1,1.2,0,\dots,0)$ and $\boldsymbol{\beta}_{log} =(1.8, -1, -0.8,0,\dots,0)$. Here, we used an imbalanced treatment assignment with a 2:1 ratio of those assigned to treatment and control, similar to that in the application study. We evaluated the model in the four scenarios outlined above, with $U_{1i}$ and $U_{2i}$  included in the data generation as necessary. }}
 
%Alternatively, our method provides direct inference on the inclusion or exclusion of a marginal mediation effect. However,    %In the second and third strategies, estimates of indirect effects are obtained by taking the mean of their posterior samples and functions thereof. Corresponding credible intervals are constructed using the appropriate quantiles of the posterior samples.  For strategy 2 and 3, we compute the posterior of marginal mediation effects for each taxa in the input. We assumed a taxa as selected if the 95\% credible interval of its corresponding mediation effect obtained from posterior samples is apart from zero. 
 
\subsection{Parameter Settings and Performance Measures} \label{ParamSettings}
%In practice, \cite{wadsworth2017integrative} suggest setting the latent inclusion indicator hyperparameters $a + b = 2$, where the prior expected proportion of active terms in the model is $a/(a+b)$. Large values of  hyperparameters $h_{c}$, $h_{\beta}$, and $h_{\kappa}$ impose vague priors on their corresponding regression coefficients. 

Simulation results were obtained by assuming uniform priors on the parameters of the Beta-Bernoulli distributions for the latent inclusion indicators in the linear and DM regression models (i.e., $  a_j= b_j= a_p= b_p= a_v= b_v= a_t= b_t = 1$). \textcolor{black}{As such, we impose no prior knowledge on whether or not covariates are active in both levels of the model as well as whether or not the balances are associated with the continuous outcome or treatment is associated with the relative abundances}. We set the scale parameters $h_{c}= h_{\beta}= h_{\kappa} = 1$, representing  weakly informative priors for the slab variances, and assumed a weakly informative prior for $\sigma^2$ in Eq.\ \eqref{stage2} (i.e., $a_0= b_0 = 1$). \textcolor{black}{ Additionally, we set the prior variances for the  intercept terms in the DM portion of the model $\sigma^2_{\alpha}= 1$ and variances for the corresponding regression coefficients for treatment and covariates associated with the relative abundances $ r^2_j = 10$ for all $j = 1, \dots, J$.} \textcolor{black}{This places a 95\% prior probability  between $\pm 1.96$ and $\pm 6.20$, respectively.} We initiated the MCMC chains at zero for all regression coefficients in both levels of the model. % and $\sigma^2 = 1$.
Each MCMC chain was run with 5000 iterations and thinned to every 10$^{th}$ iteration with a 250 iteration burn-in. Convergence was assessed by examining traceplots for $\boldsymbol{\phi}$ and $\boldsymbol{\beta}$. 

Selection performance was evaluated via sensitivity (SENS), specificity (SPEC), and Matthew's correlation coefficient (MCC), a balanced measure of the quality of binary classification \citep{Powers}. These metrics are defined as 
\begin{equation*}
 SENS = \frac{TP}{FN + TP};~~
% \end{equation*}
% \begin{equation*}
 SPEC = \frac{TN}{TN + FP}  
  \end{equation*}
 \begin{equation*}
 MCC = \frac{TP \times TN - FP \times FN}{\sqrt{(TP+FP)(TP+FN)(TN+FP)(TN+FN)}},
 \end{equation*}
where TP, TN, FP, FN represent the true positives, true negatives, false positives and false negatives based on the selection and exclusion of \textcolor{black}{relative} indirect effects. \textcolor{black}{To evaluate and compare the estimation performance for the population-level (since there are no covariates in the model) direct and overall indirect effects of the models, we calculate the average bias, mean squared error, and coverage probabilities of the equal-tail credible intervals.} Results were averaged over 50 replicated data sets for each simulated setting described in section \ref{SimulatedData}. 
%Additionally, we report the overall performance of our model, evaluated by calculating the correct selection rate (CSR) as the proportion of replicate data sets in which the model identified all active mediators and excluded all nonactive terms. 

\subsection{Methods Comparison}\label{comparison}
We compared the results of our model with the methods proposed in \citet{zhang2019} and \citet{zhang2020}, since these methods have shown superior performance in identifying \textcolor{black}{relative} indirect effects when compared to other existing methods. These models take a penalized approach to shrink regression coefficients corresponding to both the direct and indirect effects using a debiased Lasso approach \citep{ZhangandZhang}.  In both models, taxon-specific mediation effects are tested as 
\begin{eqnarray*}
    H_{0j}: \phi_1^{[j]}\beta_1^{[j]} = 0 \mbox{~~versus~~} H_{Aj}: \phi_1^{[j]}\beta_1^{[j]} \neq 0, 
\end{eqnarray*}
where the superscript $[j]$ indicates the $j^{th}$ taxon is the $1^{st}$ element in $\boldsymbol{\eta}_1$. Note that these approaches require the model to be run $J$ times for inference on each taxon, similar to CMbvs$_1$. For the method proposed by \cite{zhang2019}, which we denoted as B-H, a joint significance test is used to test the null hypothesis above. Specifically, the $p$-value for the indirect effect, $P_{joint^{[j]}}$, is set to $\max\{P_{\phi_1^{[j]}},P_{\beta_1^{[j]}}\}$,  \textcolor{black}{where $P_{\phi_1^{[j]}} = 2(1-\Phi(\mbox{abs}(\phi_1^{[j]}) / \sigma_{\phi_1^{[j]}}))$ and $P_{\beta_1^{[j]}} = 2(1-\Phi(\mbox{abs}(\beta_1^{[j]}) / \sigma_{\beta_1^{[j]}}))$, with
abs$(\cdot)$ denoting the absolute value} and $\Phi(  \cdot)$ representing the cumulative density function of a normal distribution.  For the second comparative model (CT-Lasso), \cite{zhang2020} addressed potential multiple testing issues associated with the B-H approach by proposing a closed testing-based selection procedure to calculate the $p$-value for each mediator \textcolor{black}{(see Algorithm 1 in \cite{zhang2020} for more details).}  

%\textcolor{black}{According to the first strategy, in order to investigate the marginal effect for each mediator, the model was run $J$ times, with the $j^{th}$ taxon proportion set as the  $1^{st}$ element in $\boldsymbol{\eta}_1$ for the $j^{th}$ run. } 

% shrink the column space in between 

\subsection{\textcolor{black}{Results} }
Table~\ref{table:Addi_simulation} presents the results of the simulation study across all scenarios.  With a correctly specified model (scenario 1), all methods provided excellent performance in terms of sensitivity (SENS $> 0.97$, with the exception of CMbvs$_2$ and CMbvs$_3$, which obtained SENS = 0.773 and SENS = 0.720, respectively). CMbvs$_1$, CT-Lasso, and B-H obtained the best performances overall in scenario 1, with MCC $> 0.98$. CMbvs$_2$  obtained the lowest specificity (SPEC = 0.832) among the Bayesian methods, resulting in the worst performance overall in scenario 1 (MCC = 0.357). In scenario 2, all methods were able to maintain high specificity. However, the proposed methods obtained lower sensitivity in scenario 2 relative to scenario 1. Further investigation revealed that the reduction in sensitivity for the \textcolor{black}{relative} mediation effects in the presence of model misspecifiation in the DM portion of the model resulted in poorer selection performance in the linear portion of the model (see Supplementary Table S1). We attribute this downstream reduction in performance to poorer estimation of the \textcolor{black}{relative abundances} from the DM portion of the model.  In scenario 3, CMbvs$_1$ demonstrated the best performance in terms of sensitivity, specificity, and MCC. In scenario 4, CMbvs$_1$ and CT-Lasso performed the best overall (MCC = 0.738 and MCC = 0.791, respectively). While the other methods maintained relatively high specificity, their overall performance greatly declined in the presence of unmeasured confounding due to a greater reduction in sensitivity. 

\textcolor{black}{In Table~\ref{table:est}, we report the estimation performance for the direct effect and the overall indirect effect using the proposed model (CMbvs$_1$). In scenario 1 (i.e., no unmeasured confounding), the bias for the direct effect and overall indirect effect estimated by CMbvs$_1$ was  0.662 and 1.252, respectively, the corresponding MSE was 1.255 and 4.109, respectively, and 94\% of replicates recovered the true direct effect, while all replicates recovered the true overall indirect effect in the 95\% posterior credible intervals. In scenario 3 (i.e., misspecified LM portion of the model), CMbvs$_1$ maintained similar performance as in scenario 1. However, the proposed method's estimation performance suffered in scenarios 2 and 4 (i.e., misspecification of the DM portion of the model and unmeasured confounding). These results align with the reduction in selection performance observed in these settings.  }

To further investigate model performance, we considered alternative data generation settings in scenario 1 (Table \ref{table:variation_simulation_new}). With a lower proportion of observations assigned to treatment (i.e., $P(T_i =1) = 0.25$), the methods demonstrated a reduction in sensitivity.   As expected, selection performance for all  methods decreased with a smaller sample size and larger number of compositional elements.   In the Supplementary Material, we provide additional simulation results, including more details of the selection performance for CMbvs$_1$ in both levels of the model in various scenarios (Supplementary Tables S1-S3), a comparison of all models under different data generation settings in scenario 4 (Supplementary Table S4), \textcolor{black}{as well as the acceptance probability for $\boldsymbol{\alpha}$ and $ \boldsymbol{\phi}$ across MCMC iterations (Supplementary Table S5).}  

\textcolor{black}{Table~\ref{table:J36_simulation} presents the results of the simulation study with data generated similar to the application study in the 4  scenarios outlined above. In these settings, all methods obtained lower selection performance due to the relatively small sample size, as expected. We observed that CMbvs$_1$ obtained the best overall selection performance in scenarios 1, 2, and 3 (MCC = 0.625, 0.435, and 0.564, respectively). The two comparison methods performed reasonably well when the LM portion of the model was correctly specified. However, ignoring influential covariates when modeling the multivariate count data resulted in considerably lower sensitivity and MCC for the CT-Lasso and B-H methods compared to the proposed method. } 

Taken overall, our results indicate CMbvs$_1$ as the best strategy, as it typically obtained the best overall selection performance due to high specificity across simulations. CMbvs$_2$ often gave the lowest specificity and sensitivity, resulting in the worst performance overall. CMbvs$_3$, the hybrid approach, has shown to be  more robust in small $n$ and larger $J$ settings and in the presence of unmeasured confounding. Since CMbvs$_3$ does not require $J$ fits, we recommend it in addition to CMbvs$_1$ in these select settings.   

\subsection{Sensitivity Analysis}
To investigate our model's sensitivity to prior specification, we set each of the hyperparameters to the values used in section \ref{ParamSettings} (referred to as the baseline setting) and then evaluated the effect of manipulating each term on the results obtained. Specifically, we investigated sensitivity to the prior probability of inclusion, spike-and-slab variances in the Dirichlet-multinomial model, scale parameters in the linear model, and hyperparameters for the variance of the error terms. For comparison to the baseline setting, we randomly selected a simulated data set from scenario 4 as reference and re-ran our model with all three strategies. Each MCMC algorithm was run for 5000 iterations and thinned to every $10^{th}$ iteration, with the first 250 samples as burn-in. Results of the sensitivity analysis are presented in Table \ref{table:sensitivity}. We found that with smaller prior probabilities of inclusion (i.e., 1\%, $ a_j = a_p = a_v = a_t = 0.02$ and $  b_j = b_p = b_v = b_t = 1.98$, and 10\%, $  a_j = a_p = a_v = a_t = 0.2$ and $  b_j = b_p = b_v = b_t = 1.8$) our proposed model identified fewer active mediators, as expected. We observed that  CMbvs$_1$ was relatively robust to the specification of the spike-and-slab variances in the DM portion of the model, $r_j^2$, and the scale parameters in the linear model, $h_c$, $h_{\beta}$, and $h_{\kappa}$. Lastly, we found moderate sensitivity of CMbvs$_1$ to the hyperparameter specification for the variance of the error terms in the linear model which resulted in the underselection of balances. Compared to CMbvs$_1$, CMbvs$_2$ typically obtained more false positives and CMbvs$_3$ was more sensitive to the specification of $r_j^2$.

\section{Application}
We applied the proposed joint model to a benchmark data set collected to study  the impact of sub-therapeutic antibiotic treatment on gut microbiota and body weight in early-life mice $(n=57)$ \citep{Shulfer2019}. DNA and operational taxonomic units (OTUs) were extracted with the 96-well MO BIO PowerSoil DNA Isolation Kit and   QIIME2, respectively \citep{Caporaso}. OTU counts used in this analysis were extracted on day 26 of the study, and weights of mice measured on day 116. 
Prior to analysis, we filtered out taxa with $>$90\% zero read counts and used a pseudovalue of 0.5 for zero reads when constructing the balances. Following  \cite{zhang2020}, we first analyzed the male mice samples only.  The antibiotic treatment group was assigned as exposure ($t_i$ = 1 for 23 mice) and compared with the control group ($t_i$ = 0 for 13 mice). The weights at sacrifice were treated as the outcome and standardized prior to analysis. 

\textcolor{black}{Assumptions of no unmeasured confounding can be broken up based on assumptions 4a and 4b. One (4a) is expected to hold in this application because of the randomized study design; the other (4b) is untestable. In the simulation study, we demonstrate the robustness of the proposed model under circumstances where assumption 4b is violated. In this application, unmeasured common causes of the mediator (microbiome) and outcome (body weight), would lead to violation of this assumption. Examples of such common causes could be genetics, medication use, stress, or injury. The combined strength of the associations between any of these factors and the microbiome and outcome, however, is expected to be weaker compared to the effect of diet, thus potentially limiting the potential for bias \citep{rothschild2018environment, cohen2019genetic, wen2017factors}. Microbes not considered as part of the analysis could also serve as potential mediator-outcome confounders or even treatment-induced mediator-outcome confounders, though omitted microbes typically appear in very low frequencies in the study population, which would also limit the potential for bias.}

For inference using the proposed Bayesian joint model, the MCMC algorithm was run for 15000 iterations and the chain was thinned to every $10^{th}$ iteration, with the first 1000 iterations treated as burn-in. The hyperparameters were set similar to those described in section \ref{ParamSettings}. Convergence was determined using trace plots of the regression coefficients.  %The model was run multiple times, with each taxon assigned as the first element in $\boldsymbol{\eta}_1$.
Each run took roughly 9 seconds for an overall computation time of around 5 minutes for CMbvs$_1$. Inclusion in the model was determined using the median model approach (i.e., MPPI $>$ 0.50). The results were additionally compared to the penalized approaches discussed in the simulation study.

\subsection{ \textcolor{black}{Results}}
\begin{figure}[!h]
\centering
\includegraphics[width=9cm,height=5.25in]{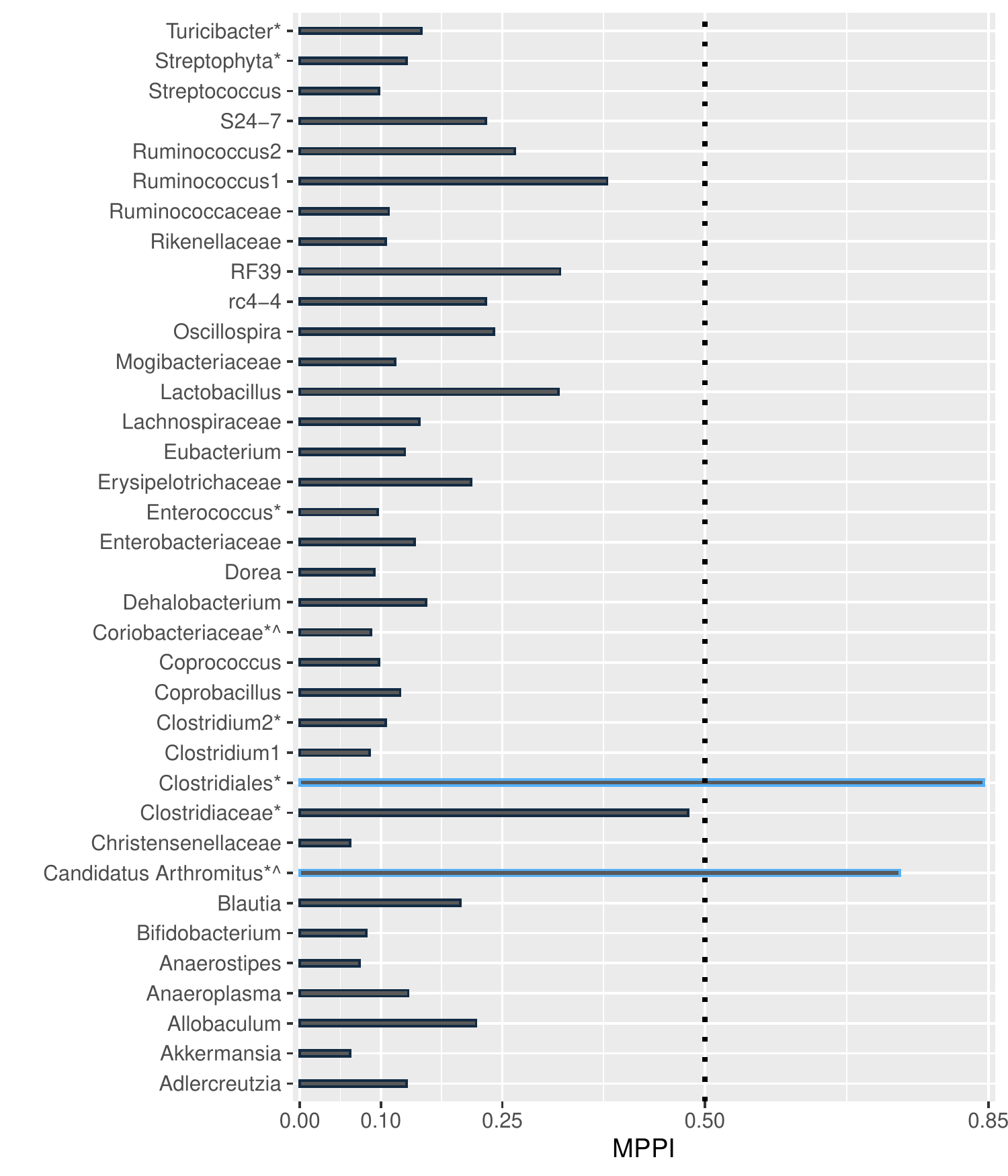}
\includegraphics[width=7cm,height=5.25in]{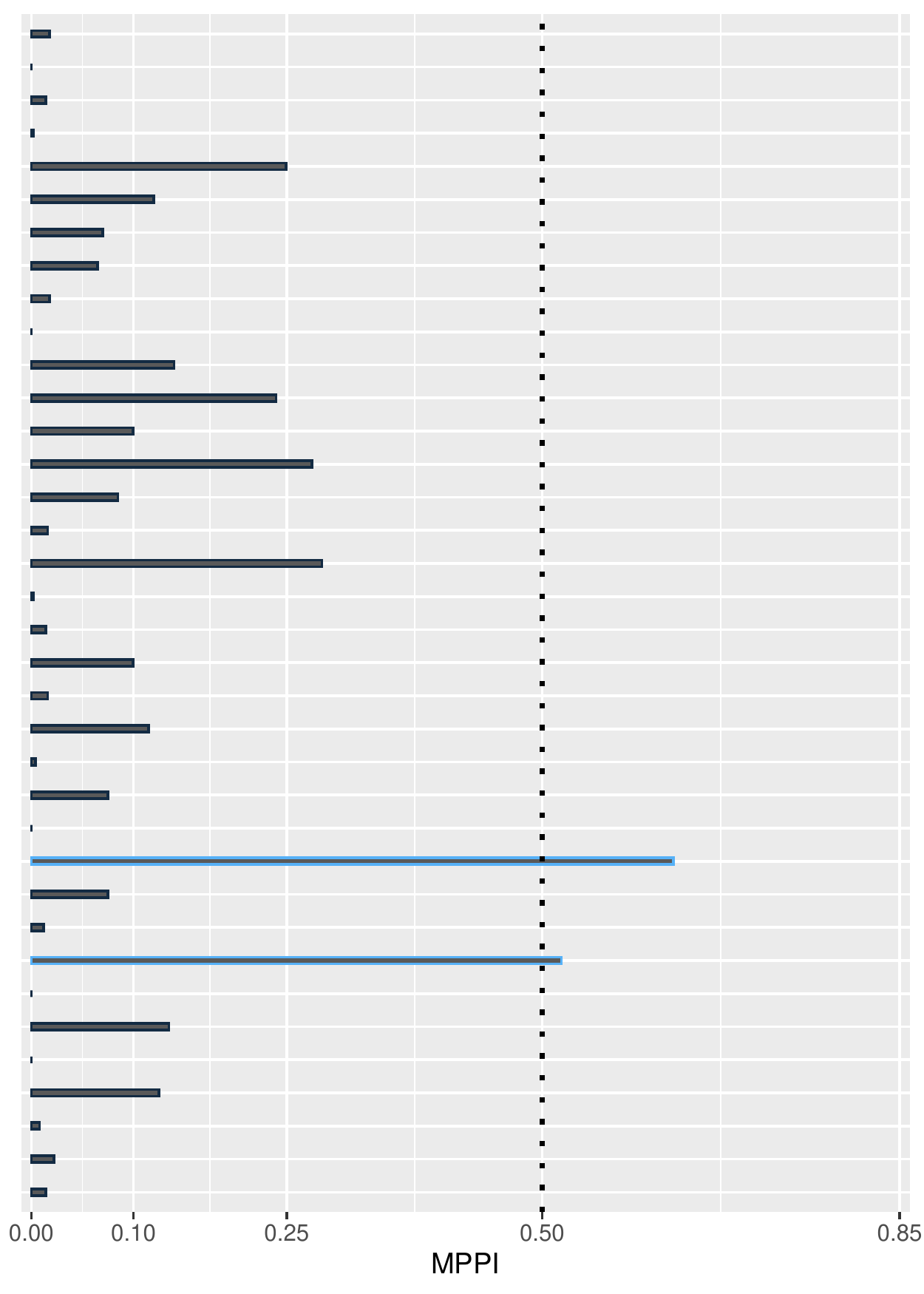}
\caption{Benchmark Study: Results from the DM portion (2a) and the outcome portion (2b) of the joint model. Marginal posterior probabilities of inclusion (MPPIs) for the corresponding $\phi_1^{[j]}$, $j=1,\dots,J$, terms (2a) and $\beta_1^{[j]}$, $j=1,\dots,J$, terms (2b), in the male mice only analysis, obtained from the $J$ runs of the model  using the CMbvs$_1$ strategy. The vertical line at 0.5 represents the inclusion threshold. Blue lines indicate selected terms.
}  
\label{Fig:MPPI_phibetanew}
\end{figure}

Plots of the corresponding MPPIs of $ \phi_1^{[j]}$ and $ \beta_1^{[j]}$ for each of the $j=1,\dots,J$ taxon in the first position of $\boldsymbol{\eta}_1$, obtained with CMbvs$_1$, are shown in Figure \ref{Fig:MPPI_phibetanew}. A 0.50 threshold on the MPPIs identified Candidatus Arthromitus and Clostridiales as potential mediators. The estimated \textcolor{black}{relative} mediation effect for Candidatus Arthromitus was $-0.033$, with a 95\% credible interval of $(-0.707, 0.004)$, and that for Clostridiales was $-0.414$ $(-1.077, 0.001)$. The estimated overall mediation effect was $-0.445$ $(-1.129,0.171)$, and the estimated direct effect of treatment on mice body weight was $0.488$  $(0.008, 1.021)$. With CMbvs$_2$, we identified Coriobacteriaceae, S24-7, and Akkermansia as potential mediators. The estimated \textcolor{black}{relative} mediation effect for Coriobacteriaceae was $0.084$, with a 95\% credible interval of $(0.015, 0.364)$,  the \textcolor{black}{relative} mediation effect for S24-7 was $0.478$ $(0.020, 1.409)$, and the \textcolor{black}{relative} mediation effect for Akkermansia was $-0.076$ $(-0.484, -0.028)$. CMbvs$_3$ did not identify any relative mediation effects. %The estimated overall mediation effect was $-0.445$ $(-1.129,0.171)$, and the estimated direct effect of treatment on mice body weight was $0.222$ $(0.000, 1.205)$. 
The B-H method also identified Candidatus Arthromitus, in addition to  Clostridiaceae, Clostridium2, Streptococcus, Coriobacteriaceae, Enterococcus, Streptophyta, and Turicibacter. The CT-Lasso method  identified Candidatus Arthromitus and Coriobacteriaceae.   

To demonstrate our approach's flexibility in accommodating and identifying potential confounders, we performed a second analysis, using the full data set, including sex in both levels of the model. We again filtered out taxa that had non-zero reads in less than 10\% of the samples, leaving 37 taxa for inference. Similar to the male-only analysis, CMbvs$_1$ selected Candidatus Arthromitus and Clostridiales as potential mediators. The \textcolor{black}{relative} mediation effect for Candidatus Arthromitus was $-0.049$ $(-0.751, 0.000)$, and that for Clostridiales was $-0.028$ $(-0.101, 0.000)$. The estimated overall mediation effect was $-0.112$ $(-0.862, 0.102)$, and the direct effect of treatment on mice weight was $0.719$ $(0.460, 0.971)$. We also identified a significant effect of sex in the outcome model with a posterior mean of $1.265$ $(1.035, 1.520)$. CMbvs$_2$ selected Erysipelotrichaceae, Streptophyta, S24-7, and Clostridiales as potential mediators with estimated \textcolor{black}{relative} mediation effects of $2.567$ $(1.872,3.241)$, $-0.137$ $(-0.703, -0.049)$, $0.425$ $(0.259, 0.637)$ and $-0.363$ $(-1.305, -0.214)$, respectively. CMbvs$_3$ also identified S24-7 and Clostridiales as potential mediators with corresponding mediation effects $0.092$ $(1.194, 2.895)$ and $0.156$ $(0.038, 0.271)$. %The estimated overall mediation effect was $-0.112$ $(-0.862, 0.102)$, and the direct effect of treatment on mice weight was $0.574$ $(0.000, 0.971)$.  
Applying the comparative models, which adjust for but do not perform selection on potential confounders, to the full data set, we observed that the B-H model identified Candidatus Arthromitus and Clostridiales, together with Streptophyta, Turicibacter and other 11 taxa. Moreover, the CT-Lasso model identified Candidatus Arthromitus in addition to Streptophyta, RF39, Clostridiaceae, Turicibacter, and Streptococcus. With the inclusion of sex in the model, we observed a reduction in the overall indirect effect  with the proposed method using CMbvs$_1$. In both the male-only and full data set analyses, we identified a ``competitive mediation effect'' as the direct effect of treatment on mice body weight and the overall mediation effect were in the opposite direction \citep{zhao2010reconsidering}. As such the microbiome acts as a suppressing effect, reducing the total effect of treatment on mice weight.

%\begin{figure}
%\centering
%\includegraphics[width=8.25cm]{plots/MPPI_beta_New}
%\caption{Benchmark study: Results from the outcome portion of the joint model. Marginal posterior probabilities of inclusion (MPPIs) for  the corresponding $\beta_1^{[j]}$, $j=1,\dots,J$, terms in the male mice only analysis obtained from the $J$ runs of the model using the CMbvs$_1$ strategy. The vertical line at 0.5 represents the inclusion threshold. Blue lines indicate selected terms. }  
%\label{Fig:MPPI_betanew}
%\end{figure}
 
\section{Discussion} 

In this work, we proposed a formulation of a Bayesian joint model for compositional data that allows for the identification, estimation, and uncertainty quantification of various causal estimands in mediation analysis. The proposed model takes advantage of sparsity-inducing priors to facilitate inference in high-dimensional compositional settings.  
Compared to existing approaches for high-dimensional compositional mediators, the proposed method employs discrete spike-and-slab priors to achieve simultaneous inference regarding the existence of direct effects, \textcolor{black}{relative} indirect effects, and overall indirect effects, in addition to potential covariates. Through simulation, we have demonstrated that our method obtains similar selection performance for \textcolor{black}{relative} mediation effects compared to existing approaches. All methods demonstrated a reduction in selection performance in the presence of unmeasured confounding and with misspecification of the linear predictor in the outcome model. The frequentist methods, CT-Lasso and B-H, were relatively robust to misspecification in the DM portion of the model. 
We have also applied our method to a benchmark data set investigating the sub-therapeutic antibiotic treatment effect on body weight in early-life mice, in which we observed a negative overall mediation effect and a positive direct effect of treatment. Overall, the proposed method identified fewer \textcolor{black}{relative} mediation effects than the alternative approaches, which was expected given the simulation results. As such, our method may favor more sparse models in practice which would result in fewer false positives but potentially more false negatives relative to the competing methods.

Using simulated data, we explored three strategies for posterior inference of \textcolor{black}{relative} mediation effects using the proposed method.  The first strategy obtained the best selection performance overall but requires refitting the model $J$ times. The second strategy, which only requires fitting the model once, demonstrated the worst selection performance overall. CMbvs$_3$, the hybrid approach, was more robust in small $n$ and larger $J$ settings and in the presence of unmeasured confounding. Based on our investigation, we recommend using the first strategy for moderate to large data sets when more sparsity is desired and additionally using CMbvs$_3$, which does not require $J$ fits, in small $n$ and larger $J$ settings and in the presence of potenital unmeasured confounding.

Using the proposed CMbvs$_1$ approach  for inference, researchers may naively cycle through each taxon without using information from the previous fit to inform the selection of the next taxon to investigate. However, in the simulation study we observed that if the $j^{th}$ taxon is associated with the outcome for $j = 2,\dots,J$, then $\beta_{j-1}$ is typically selected, regardless of the inclusion status of other terms in the outcome model. Moreover, if the \textcolor{black}{relative} mediation effect exists for a given taxon, the corresponding treatment effect in the DM portion of the model must be active. This information can be used to guide which taxon's \textcolor{black}{relative} indirect effect should be explored next, resulting in a dramatic reduction in total computation time in sparse settings. \textcolor{black}{ With CMbvs$_1$, the estimation of relative indirect effects only depends on $\beta_1$, $\alpha_1$, and $\phi_1$ (when there are no covariates in the model). However with CMbsv$_2$, relative indirect effects are dependent on a large number of estimated effects (i.e., $\beta_k$, $\alpha_k$, and $\phi_k$ for $k=1,\dots,j - 1$ for the $j^{th}$ element in $\boldsymbol{\psi}$). Thus, a major limitation of CMbsv$_2$ is that there is more opportunity for error to propagate into the estimate for $\delta_j$, and estimation performance is  highly dependent on the ordering of the taxa.} Furthermore, the three strategies were proposed for \textcolor{black}{relative} mediation effect selection when the balances are constructed using sequential binary separation. A future extension of this work would be to extend the model space to include the balance structure, in a similar spirit to the work of \cite{huang2021bayesian}, who constructed a Bayesian hierarchical model with variable selection to learn the balance structure that mediates the effect of treatment on the outcome. While this would increase computation time, it would provide simultaneous inference on the presence of any \textcolor{black}{relative} indirect effect, while fully incorporating model uncertainty. 

%This modeling approach differs from previously described methods that fix the balance structure and investigate which balances are associated with the outcome. However, their approach does not provide estimates for marginal mediation effects. 
%{\color{red}It seems this has been published %https://scholar.google.com/citations?view_op=view_citation&hl=en&user=rI_TQQEAAAAJ&cstart=20&pagesize=80&sortby=pubdate&citation_for_view=rI_TQQEAAAAJ:q-T3bmeXl90C
%}
The proposed method is designed to perform selection on each potential covariate in the model. \textcolor{black}{When the goal of the analysis is to draw inference on the treatment effect, \cite{thomas2007bayesian} and \cite{Dominici2021BMA} suggest forcing treatment into the model (i.e., not performing selection on the treatment effect). Our approach follows this setting. In simulation results not shown, we found that performing selection on the direct effect of treatment in the linear portion of the model did not affect selection performance for the \textcolor{black}{relative} mediation effects.} Also, in the formulation of our causal framework we have assumed a randomized treatment.  \textcolor{black}{Our method, however, could be extended to observational studies where treatment-outcome and treatment-mediator relationships share (measured) common causes, which can also be accounted for in the joint model. For example, in an observational study of human diet (treatment), microbiome (mediator), and obesity (outcome), we would assume that all common causes of diet and obesity are adjusted for in the model. Under this scenario, both assumptions on unmeasured confounding (4a and 4b) would be untestable. Similar sensitivity analysis for the presence of unmeasured exposure-outcome confounding can be conducted as was the case for unmeasured mediator-outcome confounding in the simulations of the current study.} Lastly, the current formulation assumes that no interaction exists between treatment and mediator, though this is not a necessary assumption for quantification of direct and indirect effects. The proposed approach could be extended to accommodate such interactions.\\

\noindent{\bf Acknowledgments:} We thank Lei Liu, Washington University in St.\ Louis,  Huilin Li and Chan Wang, New York University School of Medicine, for providing access to the benchmark data set used in the application.  \\

\noindent {\bf Data Availability Statement:} 
Code for the proposed mediation analysis method is part of the \texttt{MicroBVS} software available at https://github.com/mkoslovsky/MicroBVS. The application data that support the findings of this study were published in \cite{Shulfer2019}, to which data requests should be addressed.

\bibliographystyle{apalike}
\bibliography{MicroM}

\begin{thebibliography}{}

\bibitem[Andrews and Didelez, 2020]{andrews2020insights}
Andrews, R.~M. and Didelez, V. (2020).
\newblock Insights into the cross-world independence assumption of causal
  mediation analysis.
\newblock {\em Epidemiology}, 32(2):209--219.

\bibitem[Antonelli and Dominici, 2021]{Dominici2021BMA}
Antonelli, J. and Dominici, F. (2021).
\newblock Bayesian model averaging in causal inference.
\newblock In Tadesse, M.~G. and Vannucci, M., editors, {\em Handbook of
  Bayesian Variable Selection}. CRC Press, Boca Raton, FL.

\bibitem[Barbieri et~al., 2004]{barbieri2004optimal}
Barbieri, M.~M., Berger, J.~O., et~al. (2004).
\newblock Optimal predictive model selection.
\newblock {\em The Annals of Statistics}, 32(3):870--897.

\bibitem[Brown et~al., 1998]{brown1998multivariate}
Brown, P.~J., Vannucci, M., and Fearn, T. (1998).
\newblock Multivariate {B}ayesian variable selection and prediction.
\newblock {\em Journal of the Royal Statistical Society: Series B (Statistical
  Methodology)}, 60(3):627--641.

\bibitem[Caporaso et~al., 2010]{Caporaso}
Caporaso, J.~G., Kuczynski, J., Stombaugh, J., Bittinger, K., Bushman, F.~D.,
  Costello, E.~K., Fierer, N., Pe{\~n}a, A.~G., Goodrich, J.~K., Gordon, J.~I.,
  et~al. (2010).
\newblock {QIIME} allows analysis of high-throughput community sequencing data.
\newblock {\em Nature Methods}, 7(5):335--336.

\bibitem[Carroll et~al., 2006]{carroll2006measurement}
Carroll, R.~J., Ruppert, D., Stefanski, L.~A., and Crainiceanu, C.~M. (2006).
\newblock {\em Measurement error in nonlinear models: a modern perspective}.
\newblock Chapman and Hall/CRC.

\bibitem[Chen and Li, 2013]{chen2013variable}
Chen, J. and Li, H. (2013).
\newblock Variable selection for sparse {D}irichlet-multinomial regression with
  an application to microbiome data analysis.
\newblock {\em The Annals of Applied Statistics}, 7(1):418--442.

\bibitem[Cohen et~al., 2019]{cohen2019genetic}
Cohen, L.~J., Cho, J.~H., Gevers, D., and Chu, H. (2019).
\newblock Genetic factors and the intestinal microbiome guide development of
  microbe-based therapies for inflammatory bowel diseases.
\newblock {\em Gastroenterology}, 156(8):2174--2189.

\bibitem[Egozcue and Pawlowsky-Glahn, 2005]{egozcue2005groups}
Egozcue, J.~J. and Pawlowsky-Glahn, V. (2005).
\newblock Groups of parts and their balances in compositional data analysis.
\newblock {\em Mathematical Geology}, 37(7):795--828.

\bibitem[Egozcue et~al., 2003]{egozcue2003}
Egozcue, J.~J., Pawlowsky-Glahn, V., Mateu-Figueras, G., and Barceló-Vidal, C.
  (2003).
\newblock Isometric logratio transformations for compositional data analysis.
\newblock {\em Mathematical Geology}, 35:279--300.

\bibitem[Fewell et~al., 2007]{fewell2007impact}
Fewell, Z., Davey~Smith, G., and Sterne, J.~A. (2007).
\newblock The impact of residual and unmeasured confounding in epidemiologic
  studies: a simulation study.
\newblock {\em American Journal of Epidemiology}, 166(6):646--655.

\bibitem[George and McCulloch, 1997]{george1997approaches}
George, E.~I. and McCulloch, R.~E. (1997).
\newblock Approaches for bayesian variable selection.
\newblock {\em Statistica Sinica}, pages 339--373.

\bibitem[H{\"o}fler, 2005]{hofler2005causal}
H{\"o}fler, M. (2005).
\newblock Causal inference based on counterfactuals.
\newblock {\em BMC Medical Research Methodology}, 5(1):1--12.

\bibitem[Honkela et~al., 2001]{honkela2001nonlinear}
Honkela, A. et~al. (2001).
\newblock Nonlinear switching state-space models.
\newblock {\em Master’s thesis}.

\bibitem[Huang and Li, 2021]{huang2021bayesian}
Huang, L. and Li, H. (2021).
\newblock Bayesian balance-regression in microbiome studies using stochastic
  search.
\newblock In {\em Advances in Compositional Data Analysis}, pages 347--362.
  Springer.

\bibitem[Imai et~al., 2010]{Imai2010}
Imai, K., Keele, L., and Tingley, D. (2010).
\newblock A general approach to causal mediation analysis.
\newblock {\em Psychological methods}, 15(4):309.

\bibitem[Imai and Yamamoto, 2013]{imai2013identification}
Imai, K. and Yamamoto, T. (2013).
\newblock Identification and sensitivity analysis for multiple causal
  mechanisms: Revisiting evidence from framing experiments.
\newblock {\em Political Analysis}, 21(2):141--171.

\bibitem[Jiang et~al., 2021]{jiang2021bayesian}
Jiang, S., Xiao, G., Koh, A.~Y., Kim, J., Li, Q., and Zhan, X. (2021).
\newblock A {B}ayesian zero-inflated negative binomial regression model for the
  integrative analysis of microbiome data.
\newblock {\em Biostatistics}, 22(3):522--540.

\bibitem[Kim et~al., 2019]{kim2019bayesian}
Kim, C., Daniels, M.~J., Hogan, J.~W., Choirat, C., and Zigler, C.~M. (2019).
\newblock Bayesian methods for multiple mediators: Relating principal
  stratification and causal mediation in the analysis of power plant emission
  controls.
\newblock {\em The annals of applied statistics}, 13(3):1927.

\bibitem[Koslovsky, 2023]{koslovsky2023bayesian}
Koslovsky, M.~D. (2023).
\newblock A bayesian zero-inflated dirichlet-multinomial regression model for
  multivariate compositional count data.
\newblock {\em Biometrics}.

\bibitem[Koslovsky et~al., 2020]{koslovsky2020}
Koslovsky, M.~D., Hoffman, K.~L., Daniel, C.~R., and Vannucci, M. (2020).
\newblock A {B}ayesian model of microbiome data for simultaneous identification
  of covariate associations and prediction of phenotypic outcomes.
\newblock {\em The Annals of Applied Statistics}, 14(3):1471--1492.

\bibitem[Koslovsky and Vannucci, 2020]{koslovsky2020microbvs}
Koslovsky, M.~D. and Vannucci, M. (2020).
\newblock Micro{BVS}: {D}irichlet-tree multinomial regression models with
  {B}ayesian variable selection-an {R} package.
\newblock {\em BMC Bioinformatics}, 21(1):1--10.

\bibitem[Lin et~al., 2014]{lin2014variable}
Lin, W., Shi, P., Feng, R., and Li, H. (2014).
\newblock Variable selection in regression with compositional covariates.
\newblock {\em Biometrika}, 101(4):785--797.

\bibitem[Liu et~al., 2021]{liu2021statistical}
Liu, P., Goren, E., Morris, P., Walker, D., and Wang, C. (2021).
\newblock Statistical methods for feature identification in microbiome studies.
\newblock In {\em Statistical Analysis of Microbiome Data}, pages 175--192.
  Springer.

\bibitem[Martin-Fernandez et~al., 2000]{martin2000zero}
Martin-Fernandez, J., Barcel{\'o}-Vidal, C., and Pawlowsky-Glahn, V. (2000).
\newblock Zero replacement in compositional data sets.
\newblock In {\em Data Analysis, Classification, and Related Methods}, pages
  155--160. Springer.

\bibitem[Mart{\'\i}n-Fern{\'a}ndez et~al., 2015]{martin2015bayesian}
Mart{\'\i}n-Fern{\'a}ndez, J.-A., Hron, K., Templ, M., Filzmoser, P., and
  Palarea-Albaladejo, J. (2015).
\newblock Bayesian-multiplicative treatment of count zeros in compositional
  data sets.
\newblock {\em Statistical Modelling}, 15(2):134--158.

\bibitem[Powers, 2020]{Powers}
Powers, D. (2020).
\newblock Evaluation: From precision, recall and {F}-measure to {ROC},
  informedness, markedness \& correlation.
\newblock Technical report, Flinders University.

\bibitem[Rothschild et~al., 2018]{rothschild2018environment}
Rothschild, D., Weissbrod, O., Barkan, E., Kurilshikov, A., Korem, T., Zeevi,
  D., Costea, P.~I., Godneva, A., Kalka, I.~N., Bar, N., et~al. (2018).
\newblock Environment dominates over host genetics in shaping human gut
  microbiota.
\newblock {\em Nature}, 555(7695):210--215.

\bibitem[Rubin, 1980]{rubin1980randomization}
Rubin, D.~B. (1980).
\newblock Randomization analysis of experimental data: The fisher randomization
  test comment.
\newblock {\em Journal of the American statistical association},
  75(371):591--593.

\bibitem[Rubin, 1986]{rubin1986statistics}
Rubin, D.~B. (1986).
\newblock Statistics and causal inference: Comment: Which ifs have causal
  answers.
\newblock {\em Journal of the American Statistical Association},
  81(396):961--962.

\bibitem[Rubin, 2005]{rubin2005causal}
Rubin, D.~B. (2005).
\newblock Causal inference using potential outcomes: {D}esign, modeling,
  decisions.
\newblock {\em Journal of the American Statistical Association},
  100(469):322--331.

\bibitem[Savitsky et~al., 2011]{savitsky2011variable}
Savitsky, T., Vannucci, M., and Sha, N. (2011).
\newblock Variable selection for nonparametric {G}aussian process priors:
  {M}odels and computational strategies.
\newblock {\em Statistical Science: {A} Review Journal of the Institute of
  Mathematical Statistics}, 26(1):130--149.

\bibitem[Schulfer et~al., 2019]{Shulfer2019}
Schulfer, A.~F., Schluter, J., Zhang, Y., Brown, Q., Pathmasiri, W., McRitchie,
  S., Sumner, S., Li, H., Xavier, J.~B., and Blaser, M.~J. (2019).
\newblock The impact of early-life sub-therapeutic antibiotic treatment
  {(STAT)} on excessive weight is robust despite transfer of intestinal
  microbes.
\newblock {\em The ISME journal}, 13(5):1280--1292.

\bibitem[Sohn and Li, 2019]{sohn2017}
Sohn, M.~B. and Li, H. (2019).
\newblock Compositional mediation analysis for microbiome studies.
\newblock {\em The Annals of Applied Statistics}, 13(1):661--681.

\bibitem[Song et~al., 2021]{song2021bayesian}
Song, Y., Zhou, X., Kang, J., Aung, M.~T., Zhang, M., Zhao, W., Needham, B.~L.,
  Kardia, S.~L., Liu, Y., Meeker, J.~D., et~al. (2021).
\newblock Bayesian sparse mediation analysis with targeted penalization of
  natural indirect effects.
\newblock {\em Journal of the Royal Statistical Society: Series C (Applied
  Statistics)}.

\bibitem[Song et~al., 2020]{song2020bayesian}
Song, Y., Zhou, X., Zhang, M., Zhao, W., Liu, Y., Kardia, S.~L., Roux, A.
  V.~D., Needham, B.~L., Smith, J.~A., and Mukherjee, B. (2020).
\newblock Bayesian shrinkage estimation of high dimensional causal mediation
  effects in omics studies.
\newblock {\em Biometrics}, 76(3):700--710.

\bibitem[Tadesse et~al., 2005]{tadesse2005bayesian}
Tadesse, M.~G., Ibrahim, J.~G., Gentleman, R., Chiaretti, S., Ritz, J., and
  Foa, R. (2005).
\newblock Bayesian error-in-variable survival model for the analysis of
  genechip arrays.
\newblock {\em Biometrics}, 61(2):488--497.

\bibitem[Tadesse and Vannucci, 2021]{Vannucci2021}
Tadesse, M.~G. and Vannucci, M. (2021).
\newblock {\em Handbook of Bayesian Variable Selection}.
\newblock CRC Press, Boca Raton, FL.

\bibitem[Thomas et~al., 2007]{thomas2007bayesian}
Thomas, D.~C., Jerrett, M., Kuenzli, N., Louis, T.~A., Dominici, F., Zeger, S.,
  Schwarz, J., Burnett, R.~T., Krewski, D., and Bates, D. (2007).
\newblock Bayesian model averaging in time-series studies of air pollution and
  mortality.
\newblock {\em Journal of Toxicology and Environmental Health, Part A},
  70(3-4):311--315.

\bibitem[VanderWeele and Arah, 2011]{vanderweele2011unmeasured}
VanderWeele, T.~J. and Arah, O.~A. (2011).
\newblock Unmeasured confounding for general outcomes, treatments, and
  confounders: bias formulas for sensitivity analysis.
\newblock {\em Epidemiology}, 22(1):42.

\bibitem[VanderWeele and Vansteelandt, 2014]{VanderWeele2014}
VanderWeele, T.~J. and Vansteelandt, S. (2014).
\newblock Mediation analysis with multiple mediators.
\newblock {\em Epidemiol Methods}, 2(1):95--115.

\bibitem[Wadsworth et~al., 2017]{wadsworth2017integrative}
Wadsworth, W.~D., Argiento, R., Guindani, M., Galloway-Pena, J., Shelburne,
  S.~A., and Vannucci, M. (2017).
\newblock An integrative {B}ayesian {D}irichlet-multinomial regression model
  for the analysis of taxonomic abundances in microbiome data.
\newblock {\em BMC Bioinformatics}, 18(1):94.

\bibitem[Wang et~al., 2020]{wang2019}
Wang, C., Hu, J., Blaser, M.~J., and Li, H. (2020).
\newblock Estimating and testing the microbial causal mediation effect with
  high-dimensional and compositional microbiome data.
\newblock {\em Bioinformatics}, 36(2):347--355.

\bibitem[Wen and Duffy, 2017]{wen2017factors}
Wen, L. and Duffy, A. (2017).
\newblock Factors influencing the gut microbiota, inflammation, and type 2
  diabetes.
\newblock {\em The Journal of nutrition}, 147(7):1468S--1475S.

\bibitem[Xu et~al., 2015]{xu2015assessment}
Xu, L., Paterson, A.~D., Turpin, W., and Xu, W. (2015).
\newblock Assessment and selection of competing models for zero-inflated
  microbiome data.
\newblock {\em PloS one}, 10(7):e0129606.

\bibitem[Zhang and Zhang, 2013]{ZhangandZhang}
Zhang, C.-H. and Zhang, S.~S. (2013).
\newblock Confidence intervals for low dimensional parameters in high
  dimensional linear models.
\newblock {\em Statistical Methodology}, 76(1):317--342.

\bibitem[Zhang et~al., 2020]{zhang2020}
Zhang, H., Chen, J., Feng, Y., Wang, C., Li, H., and Liu, L. (2020).
\newblock Mediation effect selection in high-dimensional and compositional
  microbiome data.
\newblock {\em Statistics in Medicine}, 40(1):885--896.

\bibitem[Zhang et~al., 2019]{zhang2019}
Zhang, H., Chen, J., Li, Z., and Liu, L. (2019).
\newblock Testing for mediation effect with application to human microbiome
  data.
\newblock {\em Statistics in Biosciences}, 13:313--328.

\bibitem[Zhang et~al., 2018]{Zhang2018}
Zhang, J., Wei, Z., and Chen, J. (2018).
\newblock A distance-based approach for testing the mediation effect of the
  human microbiome.
\newblock {\em Bioinformatics}, 34(11):1875--1883.

\bibitem[Zhang and Yi, 2020]{zhang2020nbzimm}
Zhang, X. and Yi, N. (2020).
\newblock {NBZIMM:} negative binomial and zero-inflated mixed models, with
  application to microbiome/metagenomics data analysis.
\newblock {\em BMC Bioinformatics}, 21(1):1--19.

\bibitem[Zhang et~al., 2017]{zhang2017regression}
Zhang, Y., Zhou, H., Zhou, J., and Sun, W. (2017).
\newblock Regression models for multivariate count data.
\newblock {\em Journal of Computational and Graphical Statistics}, 26(1):1--13.

\bibitem[Zhao et~al., 2010]{zhao2010reconsidering}
Zhao, X., Lynch~Jr, J.~G., and Chen, Q. (2010).
\newblock Reconsidering baron and kenny: Myths and truths about mediation
  analysis.
\newblock {\em Journal of Consumer Research}, 37(2):197--206.

\end{thebibliography}

\clearpage
 
\begin{table}
\centering
\setlength{\tabcolsep}{2.5pt} 
\begin{tabular}{ccccccccc}
\hline
\multirow{2}{*}{Method} &  & \multicolumn{3}{c}{Scenario 1}     &  & \multicolumn{3}{c}{Scenario 2} \\ \cline{3-5} \cline{7-9} 
                        &  & SENS      & SPEC     & MCC      &  & SENS      & SPEC     & MCC      \\ \hline
CMbvs$_{1}$             &  & 0.972     & 1.000    & 0.986    &  & 0.613     & 1.000    & 0.773    \\
CMbvs$_{2}$             &  & 0.773     & 0.832    & 0.357    &  & 0.400     & 0.970    & 0.394    \\
CMbvs$_{3}$             &  & 0.720     & 0.997    & 0.809    &  & 0.383     & 0.997    & 0.567    \\
CT-Lasso                &  & 0.993     & 1.000    & 0.996    &  & 1.000     & 0.998    & 0.986    \\
B-H                     &  & 0.993     & 0.999    & 0.989    &  & 1.000     & 0.994    & 0.953    \\ \hline
                        &  & \multicolumn{3}{c}{Scenario 3} &  & \multicolumn{3}{c}{Scenario 4}     \\ \cline{3-5} \cline{7-9} 
                        &  & SENS      & SPEC     & MCC      &  & SENS      & SPEC     & MCC      \\ \hline
CMbvs$_{1}$             &  & 0.867     & 1.000    & 0.927    &  & 0.560     & 1.000    & 0.738    \\
CMbvs$_{2}$             &  & 0.713     & 0.805    & 0.294    &  & 0.326     & 0.938    & 0.233    \\
CMbvs$_{3}$             &  & 0.713     & 0.998    & 0.820    &  & 0.290     & 1.000    & 0.527    \\
CT-Lasso                &  & 0.667     & 1.000    & 0.808    &  & 0.642     & 1.000    & 0.791    \\
B-H                     &  & 0.844     & 0.991    & 0.846    &  & 0.867     & 0.944    & 0.634    \\ \hline
\end{tabular}
\caption{Simulation results for the proposed method, with three strategies for determining the \textcolor{black}{relative} mediation effects, and the comparison methods under four scenarios: (1) correctly specified model, (2) misspecification in the DM portion of the model, (3) misspecification in the linear portion of the model, and (4) unmeasured confounding. Results are averaged across 50 replicate data sets. \textbf{SENS} - sensitivity; \textbf{SPEC} - specificity; \textbf{MCC} - Matthew's correlation coefficient. }
\label{table:Addi_simulation}
\end{table}

\begin{table}[h!]
\centering
\begin{tabular}{cccccccc}
\hline
\multirow{2}{*}{Scenario} & \multicolumn{3}{c}{Direct Effect} &  & \multicolumn{3}{c}{Overall Indirect Effect} \\ \cline{2-4} \cline{6-8} 
                          & Bias      & MSE         & COV     &  & Bias          & MSE            & COV        \\ \hline
1                         & 0.662     & 1.255       & 0.94    &  & 1.252         & 4.108          & 1.00       \\
2                         & 6.600     & 85.176      & 0.65    &  & 11.153        & 215.058        & 0.78       \\
3                         & 0.805     & 1.140       & 0.86    &  & 1.611         & 6.022          & 0.94       \\
4                         & 8.668     & 101.709     & 0.54    &  & 14.658        & 293.547        & 0.62       \\ \hline
\end{tabular}
\caption{\textcolor{black}{Simulation results for estimating the direct and overall indirect effects using CMbvs$_1$ with $n=200$ and $J=50$ averaged over 50 simulations. \textbf{Bias} - the difference between the posterior mean effect and the true effect. \textbf{MSE} - the mean squared error between the posterior mean effect and the true  effect. \textbf{Coverage} - the proportion of replications in which the 95\% posterior credible interval of the effect included the true value.}}
\label{table:est}
\end{table}

\clearpage

 \begin{table}
\centering
\setlength{\tabcolsep}{1.5pt}
% \usepackage{multirow} 
\begin{tabular}{ccccccccccccc}
\hline
\multirow{2}{*}{Method} &  & \multicolumn{3}{c}{$P(T_i=1) = 0.25$}                                         &  & \multicolumn{3}{c}{$n=50$, $J=50$}   &  & \multicolumn{3}{c}{$n=50$, $J=100$}                                     \\ \cline{3-5} \cline{7-9} \cline{11-13}
                        &  & \multicolumn{1}{c}{SENS} & \multicolumn{1}{c}{SPEC} & \multicolumn{1}{c}{MCC} &  & \multicolumn{1}{c}{SENS} & \multicolumn{1}{c}{SPEC} & \multicolumn{1}{c}{MCC} &  & \multicolumn{1}{c}{SENS} & \multicolumn{1}{c}{SPEC} & \multicolumn{1}{c}{MCC}\\ \hline
CMbvs$_{1}$             &  & 0.833                    & 1.000                    & 0.908                   &  & 0.433                    & 1.000                    & 0.646          &  & 0.247                    & 0.988                    & 0.454             \\
CMbvs$_{2}$             &  & 0.673                    & 0.970                    & 0.603                   &  & 0.380                    & 0.962                    & 0.345       &  & 0.460                    & 0.935                    & 0.331             \\
CMbvs$_{3}$             &  & 0.633                    & 0.998                    & 0.758                   &  & 0.340                    & 0.999                    & 0.565         & & 0.420 & 0.998 & 0.615         \\
CT-Lasso                &  & 0.800                    & 1.000                    & 0.889                   &  & 0.233                    & 0.999                    & 0.454        & & 0.280 & 0.999 & 0.502           \\
B-H                     &  & 0.947                   & 1.000                    & 0.971                  &  & 0.347                    & 1.000                    & 0.577  & & 0.233 & 1.000  & 0.471               \\ \hline
                        
\end{tabular}
\caption{Simulation results for the proposed method, with three strategies for determining the \textcolor{black}{relative} mediation effects, and the comparison methods in scenario 1 with various data structures. Results are averaged across 50 replicate data sets. \textbf{SENS} - sensitivity; \textbf{SPEC} - specificity; \textbf{MCC} - Matthew's correlation coefficient.}
\label{table:variation_simulation_new}
\end{table}

\begin{table}
\centering
\setlength{\tabcolsep}{2.5pt} 
\begin{tabular}{ccccccccc}
\hline
\multirow{2}{*}{Method} &  & \multicolumn{3}{c}{Scenario 1}     &  & \multicolumn{3}{c}{Scenario 2} \\ \cline{3-5} \cline{7-9} 
                        &  & SENS      & SPEC     & MCC      &  & SENS      & SPEC     & MCC      \\ \hline
CMbvs1 &  & 0.422     & 0.999 & 0.625  & & 0.233     & 0.997 & 0.435 \\
CMbvs2 &   & 0.400       & 0.838 & 0.171 & 
 &  0.267     & 0.974 & 0.317 \\
CMbvs3  &  & 0.278     & 0.996 & 0.471 & & 0.267     & 0.999 & 0.380  \\
CT-Lasso &  &  0.289       &  1.000     &   0.520  &   &  0.060       &   1.000    &   0.175    \\
B-H      &  &   0.353      &   0.006    &   0.544  &   &   0.087   &    1.000   &   0.287    \\
\hline
                        &  & \multicolumn{3}{c}{Scenario 3} &  & \multicolumn{3}{c}{Scenario 4}     \\ \cline{3-5} \cline{7-9} 
                        &  & SENS      & SPEC     & MCC      &  & SENS      & SPEC     & MCC      \\ \hline
CMbvs1 &  & 0.333     & 1.000     & 0.564 & & 0.167     & 0.995 & 0.356 \\
CMbvs2 &  & 0.456     & 0.877 & 0.257 & & 0.244     & 0.974 & 0.290  \\
CMbvs3  &  & 0.322     & 0.995 & 0.506 & & 0.156     & 1.000     & 0.380  \\
CT-Lasso &  &    0.200     &    1.000   &    0.431   &  &   0.000      &    1.000   &    0.000   \\
B-H      &  &    0.260     &   0.998    & 0.470   &  &  0.033        &    1.000   &   0.175    \\
\hline
\end{tabular}
\caption{\textcolor{black}{Simulation results for the proposed method on data simulated similar in structure to the application data using the three strategies for determining the \textcolor{black}{relative} mediation effects and the comparison methods under four scenarios: (1) correctly specified model, (2) misspecification in the DM portion of the model, (3) misspecification in the linear portion of the model, and (4) unmeasured confounding. Results are averaged across 50 replicate data sets. \textbf{SENS} - sensitivity; \textbf{SPEC} - specificity; \textbf{MCC} - Matthew's correlation coefficient. }}
\label{table:J36_simulation}
\end{table}

\clearpage
 
\begin{table}
\centering
\setlength{\tabcolsep}{1.7pt}
\begin{tabular}{ccccccccccccc}
\hline
\multirow{2}{*}{Method} &  & \multicolumn{3}{c}{Baseline}                                                  &  & \multicolumn{3}{c}{PrPI = 1\%}                                                &                      & \multicolumn{3}{c}{PrPI = 10\%}                                               \\ \cline{3-5} \cline{7-9} \cline{11-13} 
                        &  & \multicolumn{1}{c}{SENS} & \multicolumn{1}{c}{SPEC} & \multicolumn{1}{c}{MCC} &  & \multicolumn{1}{c}{SENS} & \multicolumn{1}{c}{SPEC} & \multicolumn{1}{c}{MCC} &                      & \multicolumn{1}{c}{SENS} & \multicolumn{1}{c}{SPEC} & \multicolumn{1}{c}{MCC} \\ \hline
CMbvs$_{1}$             &  & 0.667                    & 1.000                    & 0.808                   &  & 0.667                    & 1.000                    & 0.808                   &                      & 0.333                    & 1.000                    & 0.565                   \\
CMbvs$_{2}$             &  & 0.667                    & 0.979                    & 0.645                   &  & 1.000                    & 0.723                    & 0.368                   &                      & 0.333                    & 0.894                    & 0.166                   \\
CMbvs$_{3}$             &  & 0.667                    & 1.000                    & 0.808                   &  & 0.667                    & 0.979                    & 0.645                   &                      & 0.667                    & 1.000                    & 0.808                   \\ \hline
                        &  & \multicolumn{3}{c}{$r_j^2 = 5$}                                               &  & \multicolumn{3}{c}{$r_j^2 = 20$}                                              &                      & \multicolumn{3}{c}{$h_{c,\beta,\kappa} = 5$}                                  \\ \cline{3-5} \cline{7-9} \cline{11-13} 
                        &  & \multicolumn{1}{c}{SENS} & \multicolumn{1}{c}{SPEC} & \multicolumn{1}{c}{MCC} &  & \multicolumn{1}{c}{SENS} & \multicolumn{1}{c}{SPEC} & \multicolumn{1}{c}{MCC} &                      & \multicolumn{1}{c}{SENS} & \multicolumn{1}{c}{SPEC} & \multicolumn{1}{c}{MCC} \\ \hline
CMbvs$_{1}$             &  & 1.000                    & 1.000                    & 1.000                   &  & 0.667                    & 1.000                    & 0.808                   &                      & 0.667                    & 1.000                    & 0.808                   \\
CMbvs$_{2}$             &  & 0.333                    & 0.957                    & 0.291                   &  & 0.667                    & 0.851                    & 0.320                   &                      & 0.667                    & 0.830                   & 0.294                   \\
CMbvs$_{3}$             &  & 0.333                    & 1.000                    & 0.565                   &  & 0.333                    & 0.979                    & 0.378                   &                      & 1.000                    & 1.000                    & 1.000                   \\ \hline
                        &  & \multicolumn{3}{c}{$h_{c,\beta,\kappa} = 20$}                                 &  & \multicolumn{3}{c}{$a_0 = b_0 = 0.1$}                                         & \multicolumn{1}{c}{} & \multicolumn{3}{c}{$a_0 = b_0 = 10$}                                          \\ \cline{3-5} \cline{7-9} \cline{11-13} 
                        &  & \multicolumn{1}{c}{SENS} & \multicolumn{1}{c}{SPEC} & \multicolumn{1}{c}{MCC} &  & \multicolumn{1}{c}{SENS} & \multicolumn{1}{c}{SPEC} & \multicolumn{1}{c}{MCC} &                      & \multicolumn{1}{c}{SENS} & \multicolumn{1}{c}{SPEC} & \multicolumn{1}{c}{MCC} \\ \hline
CMbvs$_{1}$             &  & 0.667                    & 1.000                    & 0.808                   &  & 0.333                    & 1.000                    & 0.565                   &                      & 1.000                    & 1.000                    & 1.000                   \\
CMbvs$_{2}$             &  & 0.667                    &0.979                    & 0.645                   &  & 0.000                    & 1.000                    & 0.000                   &                      & 0.667                    & 0.787                    & 0.069                   \\
CMbvs$_{3}$             &  & 0.667                    & 1.000                    & 0.808                   &  & 0.333                    & 1.000                    & 0.565                  &                      & 1.000                    & 0.979                    & 0.857                   \\ \hline
\end{tabular}
\caption{Results of the sensitivity analysis for the proposed method with three selection strategies on data simulated from scenario 4. Baseline settings refers to the model fit with hyperparameter settings from the simulation study. \textbf{PrPI} - prior probability of inclusion; \textbf{SENS} - sensitivity; \textbf{SPEC} - specificity; \textbf{MCC} - Matthew's correlation coefficient.  }
\label{table:sensitivity}
\end{table}

\end{document}